\documentclass{article}

\usepackage{arxiv}

\usepackage[utf8]{inputenc} 
\usepackage[T1]{fontenc}    
\usepackage[hidelinks]{hyperref}       
\usepackage{url}            
\usepackage{booktabs}       
\usepackage{amsfonts}       
\usepackage{nicefrac}       
\usepackage{microtype}      
\usepackage{lipsum}
\usepackage[table]{xcolor} 
\usepackage{graphicx}
\usepackage{subcaption} 
\usepackage{multirow}
\usepackage{makecell}
\usepackage{acronym}
\usepackage{tabularx}
\usepackage{amsmath}
\usepackage{natbib}

\definecolor{fixsac}{RGB}{230,240,255}
\definecolor{hda}{RGB}{240,230,255}
\definecolor{fused}{RGB}{255,240,230}

\graphicspath{ {./images/} }

\acrodef{AFT}{Average Fixation Time}
\acrodef{AIC}{Akaike Information Criterion}
\acrodef{AUC}{Area Under the Receiving Operating Characteristic Curve}
\acrodef{BDI}{Beck Depression Inventory}
\acrodef{BIC}{Bayesian Information Criterion}
\acrodef{CBEM}{Content-Based Ensemble Models}
\acrodef{DT}{Decision Tree}
\acrodef{EEG}{Electroencephalogram}
\acrodef{ECG}{Electrocardiogram}
\acrodef{EF}{Experimental Framework}
\acrodef{EM}{Expectation-Maximization}
\acrodef{EoS}{Entropy of States}
\acrodef{FFS}{Forward Feature Selection}
\acrodef{FO}{Fractional Occupancy}
\acrodef{GMM}{Gaussian Mixture Model}
\acrodef{HC}{Healthy Control}
\acrodef{HDA}{High-Density Area}
\acrodef{HGUGM}{Hospital General Universitario Gregorio Marañón}
\acrodef{HUF}{Hospital Universitario de Fuenlabrada}
\acrodef{HMM}{Hidden Markov Model}
\acrodef{LD}{Longest Diagonal}
\acrodef{LR}{Logistic Regression}
\acrodef{MC}{Markov Chain}
\acrodef{MDS-UPDRS}{Movement Disorder Society - Unified Parkinson’s Disease Rating Scale}
\acrodef{MLT}{Mean Lifetime}
\acrodef{MIL}{Mean Interval Length}
\acrodef{MoCA}{Montreal Cognitive Assessment}
\acrodef{MoE}{Mixture of Experts}
\acrodef{NB}{Naïve Bayes}
\acrodef{PD}{Parkinson's Disease}
\acrodef{RBF}{Radial Basis Function}
\acrodef{RF}{Random Forest}
\acrodef{ROI}{Region of Interest}
\acrodef{SVM}{Support Vector Machines}
\acrodef{TF}{Total Fixations}
\acrodef{TS}{Total Saccades}
\acrodef{TSA}{Total Scanned Area}
\acrodef{TSE}{Total Saccadic Excursion}
\acrodef{yHC}{Young Healthy Control}

\title{Automatic Screening of Parkinson's Disease From Visual Explorations}

\author{
 Maria F. Alcala-Durand \\
  Escuela Técnica Superior de Ingenieros de Telecomunicación\\
  Universidad Politécnica de Madrid\\
  \texttt{maria.adurand@upm.es} \\
    \And
 J. Camilo Puerta-Acevedo \\
  Escuela Técnica Superior de Ingenieros de Telecomunicación\\
  Universidad Politécnica de Madrid\\
 \texttt{juancamilo.puerta@upm.es}  \\
  \And
 Julián D. Arias-Londoño \\
  Escuela Técnica Superior de Ingenieros de Telecomunicación\\
  Universidad Politécnica de Madrid\\
  \texttt{julian.arias@upm.es}  \\
  \And
 Juan I. Godino-Llorente \\
  Escuela Técnica Superior de Ingenieros de Telecomunicación\\
  Universidad Politécnica de Madrid\\
  \texttt{ignacio.godino@upm.es} \\
}

\begin{document}
\maketitle
\begin{abstract}

Eye movements can reveal early signs of neurodegeneration, including those associated with Parkinson\textquotesingle s Disease (PD). This work investigates the utility of a set of gaze-based features for the automatic screening of PD from different visual exploration tasks. For this purpose, a novel methodology is introduced, combining classic fixation/saccade oculomotor features (e.g., saccade count, fixation duration, scanned area) with features derived from gaze clusters (i.e., regions with a considerable accumulation of fixations). These features are automatically extracted from six exploration tests and evaluated using different machine learning classifiers. A Mixture of Experts ensemble is used to integrate outputs across tests and both eyes. Results show that ensemble models outperform individual classifiers, achieving an \ac{AUC} of 0.95 on a held-out test set. The findings support visual exploration as a non-invasive tool for early automatic screening of PD.

\end{abstract}


\section{Introduction}

Video-based eye tracking has become a valuable tool for quantifying oculomotor behaviour in clinical research. It has been applied across a variety of conditions using tasks such as smooth pursuit, anti-saccades, and visual exploration, demonstrating utility in identifying markers of neurological and psychiatric disorders \cite{ reiner2023oculometric, matsumoto2011small, bek2020measuring, tsang2016eye, armstrong2012eye}. These methods have been applied to evaluate clinical signs of Alzheimer\textquotesingle s disease \cite{davis2020eye, boz2023examination}, epilepsy \cite{metternich2022eye}, aphasia \cite{ashaie2020eye}, and autism spectrum disorder \cite{wang2015atypical}, reporting changes in gaze dynamics due to the specific health condition, such as reduced scanpath length, gaze rigidity, or altered saliency responses.

Eye tracking using videoculographic techniques is also a valuable tool in the clinical research of \ac{PD}. Although \ac{PD} is clinically characterised by coarse motor impairments (such as tremor, rigidity, and bradykinesia, which are associated with dysfunction in the primary motor cortex and basal ganglia), patients also present impairments in fine motor functions, including oculomotor control and speech production \cite{antoniades2024eye, kassavetis2022eye, rodriguez2019eye, leigh2020abnormal}.  


Several studies have documented specific eye movement abnormalities in patients with \ac{PD}. These studies are typically carried out using controlled oculomotor tasks such as smooth pursuit or prosaccade tests. Patients often exhibit increased saccadic latency, hypometric saccades, reduced smooth pursuit gain, and impaired fixation stability \cite{reiner2023oculometric, antoniades2024eye, kassavetis2022eye, leigh2020abnormal, wong2020prolonged, dietz2011emotion}. Nonetheless, other features are found relevant in uncontrolled exploration tasks, where participants observe complex images without explicit instructions. Among others, patients often exhibit longer fixation durations, shorter saccadic amplitudes, and smaller scanned areas \cite{matsumoto2011small, wong2020prolonged}. These measures have also demonstrated correlation with impairments in memory and verbal fluency in \ac{PD} patients without dementia \cite{wong2018eye}. Furthermore, emotional content also influences the gaze behaviour of \ac{PD} patients in exploration tasks. When viewing emotionally charged images, patients tend to exhibit reduced scanpaths and fewer fixations \cite{dietz2011emotion}. Related findings have been observed in certain mood disorders such as depression, where individuals allocate more attention to negative stimuli than to positive ones \cite{takemoto2023depression, arndt2014eye, russell2015eye}.



Beyond the aforementioned standard metrics, modelling scanpath dynamics has emerged as a method for capturing the temporal structure of gaze during free visual exploration tasks \cite{coutrot2018scanpath}. This method is based on modelling the scanpath using a \ac{HMM} \cite{rabiner1989tutorial}, in which the underlying system is assumed to be a Markov process with a finite number of unobserved (hidden) states and associated state transitions following a probability distribution. In this context, a state is loosely defined as a region with a considerable accumulation of data points, akin to a cluster. 

Alternative studies have applied a similar \ac{HMM}-based modelling strategy to other application domains, such as brain activity detection from magnetoencephalographic  (MEG) recordings \cite{tibon2021transient} or functional magnetic resonance (rs-fMRI) \cite{bustamante2023classification}, and gait recognition using data collected from inertial sensors \cite{khorasani2014hmm}. Besides, literature also reports the applications of \acp{HMM} combined with entropy-based measures to model the temporal structure and variability of transitions between latent states, such as in voice signal processing contexts \cite{arias2015entropies}. These works illustrate the broad applicability of state-based temporal modelling frameworks, though none focus on eye movement sequences.

In this context, descriptors such as \ac{FO}, \ac{MLT}, \ac{MIL}, and \ac{EoS}, have been proposed to characterise the temporal dynamics in latent state sequences. These features were proposed to extract relevant information from an \ac{HMM} in applications such as MEG-based cognitive state modelling, brain activity decoding, or the screening of voice disorders  \cite{tibon2021transient, heideman2020dissecting, zhu2023mibfm, arias2015entropies}. While such features offer more nuanced temporal insight than classic static fixation/saccade summaries, existing approaches typically require manual annotations to associate each latent state with a meaningful \ac{ROI}, a step that introduces subjectivity and reduces scalability.

In parallel, prior works have explored temporal modelling of behavioural sequences using state-based techniques such as \acp{GMM}, not only in gaze tracking \cite{coutrot2017scanpath} but also in other application domains like gait analysis in \ac{PD} \cite{khorasani2014hmm}, brain activity detection \cite{tibon2021transient}, and speech processing \cite{arias2010automatic, arias2015entropies}. These approaches extract descriptors from the dynamics of latent state transitions, offering more nuanced insights than static metrics. However, few of these studies focus on visual explorations, and none provide an automated, generalisable approach for mapping latent gaze states to different \acp{ROI} in clinical contexts.

Despite the potential of the approaches mentioned above, no study has yet developed automatically extracted state or \ac{HDA}-based visual exploration descriptors, leaving a significant methodological gap at the intersection of unguided gaze behaviour and the screening of neurodegenerative diseases (specifically \ac{PD}). 

This study addresses this gap by introducing a novel methodology that not only extracts fixation/saccade-based metrics but also automates the identification of meaningful Regions of Interest \acp{ROI} through \acp{HDA}, which are derived using an unsupervised method based on a \acp{GMM} fitted to gaze data from visual exploration tasks. 

Unlike prior approaches that require manual annotation of \acp{ROI}, this method infers latent gaze clusters directly from data, enabling scalable, objective, and reproducible computation of spatiotemporal descriptors. For this purpose, six structured images were shown to participants from three cohorts, namely: \ac{PD} patients, age-matched \ac{HC}, and \ac{yHC}. These features were used to train several machine learning classifiers. Additionally, a model based on a \ac{MoE} was employed to integrate scores across tasks and eyes, thereby producing a final patient-level score. To the best of our knowledge, this is the first study to apply fully automated \ac{HDA}-based gaze features (also in combination with fixation/saccade-based features) extracted from exploration tasks for the screening of \ac{PD} using machine learning techniques.

The rest of the paper is organised as follows: Section \ref{sec:materials} presents the corpus of videoculographic signals used; Section \ref{sec:methods} describes the methodology followed; Section \ref{sec:results} reports the results obtained; and Section \ref{sec:conclussion} draws conclusions and proposes future lines of research. 

\section{Materials}
\label{sec:materials}

This section presents the characteristics of the corpus collected: the exploration tasks, the inclusion criteria, and the population recorded. The preprocessing methods applied to the corpus are also described in this section.   

Over the course of two years, and with the help of expert physicians, several oculographic tasks were recorded. The population includes three cohorts: pre-screened \ac{PD} patients, age-matched \ac{HC}, and \ac{yHC}. Data were collected at two hospitals of the Madrid community, Spain: the \ac{HUF} and the \ac{HGUGM}. 

\paragraph{Participants.}

A total of 52 participants with \ac{PD} and 48 \ac{HC} took part in the study. The average age for the \ac{PD} cohort was 63.8 (range 44-84 years), and for the age-matched control cohort was 64.26 years (range 46-80 years). To differentiate between effects attributable to normal cognitive ageing and those specific to \ac{PD}, an additional cohort of 12 \ac{yHC} of average age 23.30 years (range 23-30) was also included.

All participants underwent clinical evaluations, and their medical histories were meticulously recorded. This included details such as the age of the onset of \ac{PD}, duration of the disease, symptoms, and any associated complications. Patients in the \ac{PD} cohort had less than 5 years of evolution. Assessments for the \ac{PD} cohort included the \ac{MDS-UPDRS} Part III and the \ac{MoCA} tests. The mean \ac{MDS-UPDRS} scores were 16.88 for the \ac{PD} cohort and 1.5 for the \ac{HC} cohort. 
The mean \ac{MoCA} scores were 25.22 for \ac{PD}, 26.74 for \ac{HC}, and 28.78 for \ac{yHC}. 

Patients with \ac{PD} were assessed in the ON medication state, having taken their usual morning dose according to their regular schedule. Recordings were typically conducted before noon.

\paragraph{Recording.}

Eye movements were tracked using an infrared, video-based binocular EyeLink\textsuperscript{®} 1000 Plus eyetracker system with a sampling rate of 1 kHz. The setup involved two computers, one dedicated to controlling the eyetracker and another to presenting visual stimuli. The stimuli (i.e., images) were displayed on a 1920$\times$1080~px LED monitor placed 60 cm in front of the participant. The illumination and acoustic conditions of the room were carefully controlled and consistently maintained for all recordings. 

Participants were comfortably seated with their heads stabilised on a chin rest to minimise movement and ensure consistent measurements. The distance between the upper knob of the eye tracking camera and the front of the chin rest was 50 cm. Before each recording session, the eye-tracking system was calibrated using a 9-point grid spanning the area where the targets appeared, ensuring accurate gaze tracking.

\begin{figure} 
    \centering
    \includegraphics[width=0.6\textwidth]{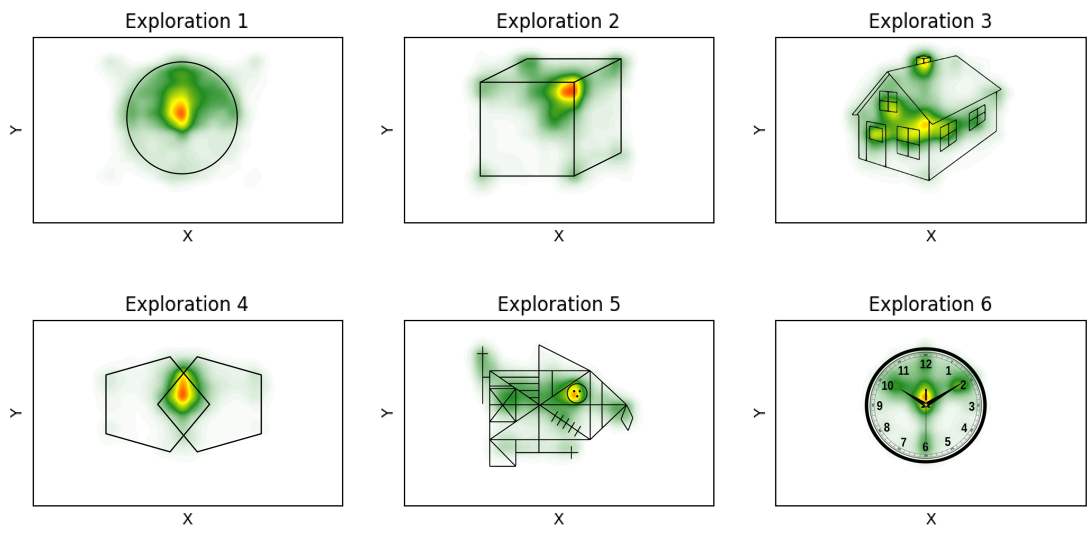}
    \caption{Exploration patterns of all participants overlaid on the shown images. Six images were presented to the patients: a cube, a house, a pair of intersecting pentagons, the Rey-Osterrieth figure, and a clock.}
    \label{fig:all_explorations}
\end{figure}

During the recording procedure, participants visually explored six different images presented sequentially and in a fixed sequence. The images, in order, are (Fig.~\ref{fig:all_explorations}): a circle, a cube, a house, a pair of intersecting pentagons, the Rey-Osterrieth complex figure \cite{osterrieth1944test}, and a clock with numbers and hands. These images have an aspect ratio of 5:4. Patients were instructed to visually explore each image in its entirety within a predetermined time frame of 15~s. Throughout this process, the system continuously recorded eye positions. For brevity, each image is referred to in the following as \textit{Expl.~1} through \textit{Expl.~6}, corresponding to the aforementioned presentation order. 

Figure~\ref{fig:all_explorations} shows the average exploration patterns of all participants (regardless of cohort) as density overlaid on the corresponding stimuli, which offers a visual summary of where gaze activity was most concentrated during the task.

\subsection{Data Preprocessing}

During recording, each of the approximately 15,000 data points per test (i.e., 15 seconds at 1 kHz) is automatically labelled as belonging to a fixation, saccade, or blink. 

A blink is detected when the pupil size is very small, missing or severely distorted by eyelid occlusion. Blink segments include unreliable velocity and position data, and must be discarded to avoid artifacts. In the raw signal, blinks often manifest as abrupt, upward deflections resembling high-velocity saccades, accompanied by distortions resulting from partial or complete occlusion of the pupil. Thus, all samples tagged as part of a blink event were excluded from the analysis. Given that the features extracted in this study are not reliant on temporal continuity of the eye movement sequence, this point-wise removal is both appropriate and methodologically sound. On average, this procedure removes 6.3\% of each recording. Blinks are automatically detected by the online parsing system provided by the eyetracker.

\begin{figure}[ht]
    \centering
    \includegraphics[width=0.6\textwidth]{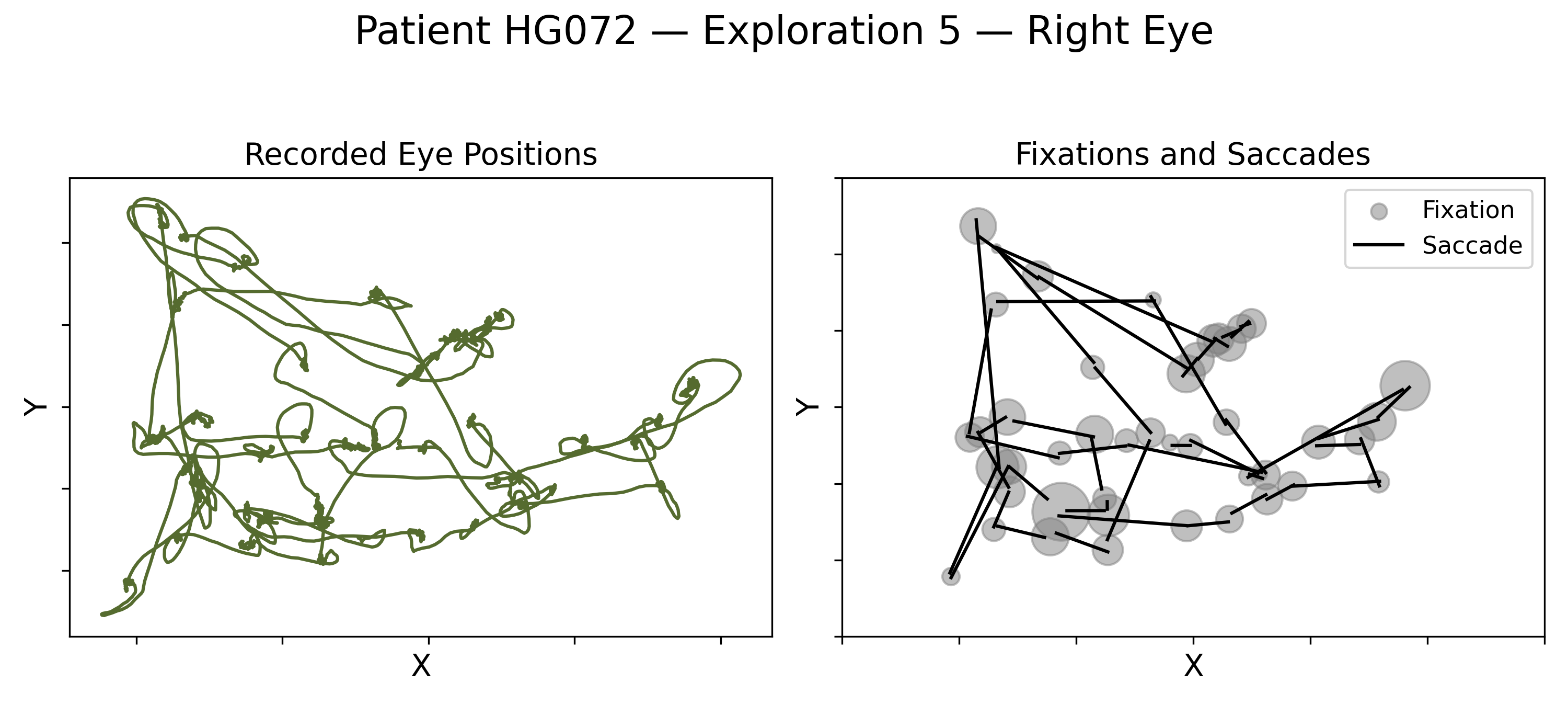}
    \caption{Recorded eye movement trace (left) and automatically detected fixations and saccades (right) for patient HG072 during one visual exploration task. 
    }
    \label{fig:eye_trace_and_events}
\end{figure}

Fixations and saccades are also automatically detected by the online parsing system provided by the eyetracker, which uses a method based on the velocity and acceleration of gaze \cite{srresearch2017eyelink}. Figure~\ref{fig:eye_trace_and_events} shows an example of these detected events for a single patient. The left panel displays the spatial trace of recorded gaze positions over time. The right panel visualises the resulting segmentation into saccades (black lines) and fixations (grey circles), with circle size proportional to the fixation duration.

Out of the 1,416 available observations (spanning six visual explorations, two eyes per participant, and 118 participants), 69 were excluded due to data quality concerns. Exclusion criteria fell into three categories: (1) insufficient data, including cases with total measured time below the 1st percentile or excessive blinking above the 99th percentile; (2) blink-related data loss, such as a high blink-to-event ratio, excessive blink duration, or prolonged blink time—all exceeding the 99th percentile; and, (3) inter-eye dissimilarity, specifically extreme temporal lag in the cross-correlation of $x$- or $y$-coordinates, based on blink-free data. These empirically defined thresholds ensured robust and reliable input for subsequent analyses.

Gaze coordinates, $\mathbf{x}_i=(x_i, y_i) \in \mathbb{R}^2$, were linearly normalised to the $[0, 1]$ range based on the eyetracker\textquotesingle s default output space, preserving the original aspect ratio. All analyses were performed using these normalised data.

\section{Methods}
\label{sec:methods}

\begin{figure} 
    \centering  \includegraphics[width=1\textwidth]{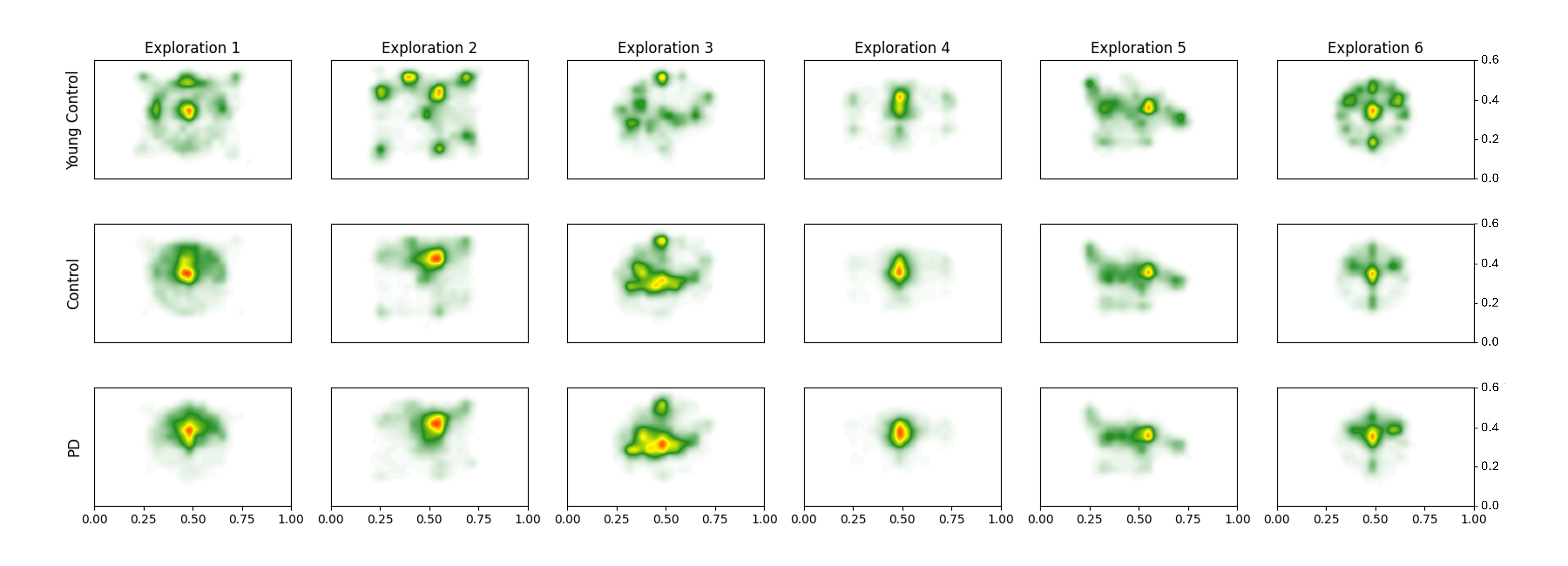}
    \caption{Prototypical examples of the exploration patterns for the three different cohorts available, per exploration.}
    \label{fig:all_explorations_grouped}
\end{figure}

A perceptual visualisation of the gaze density distribution suggests tangible differences between cohorts in their respective strategies for exploring images. 
Roughly speaking, \ac{PD} patients tend to be more rigid while exploring the images, thus covering less area and staying longer in certain regions. This behaviour is detailed in Fig.~\ref{fig:all_explorations_grouped}, and represents the basis for the hypotheses drawn in this work and for the methods proposed to model such behaviour.

This section describes the methods used to characterise the explorations, grouping them into two sets: standard static fixation/saccade metrics, and \ac{HDA}-based features. Besides, it presents the classification machines used, the \ac{MoE} strategies considered, and the experimental framework followed. 

\begin{figure}[ht]
    \centering
    \includegraphics[scale=0.6]{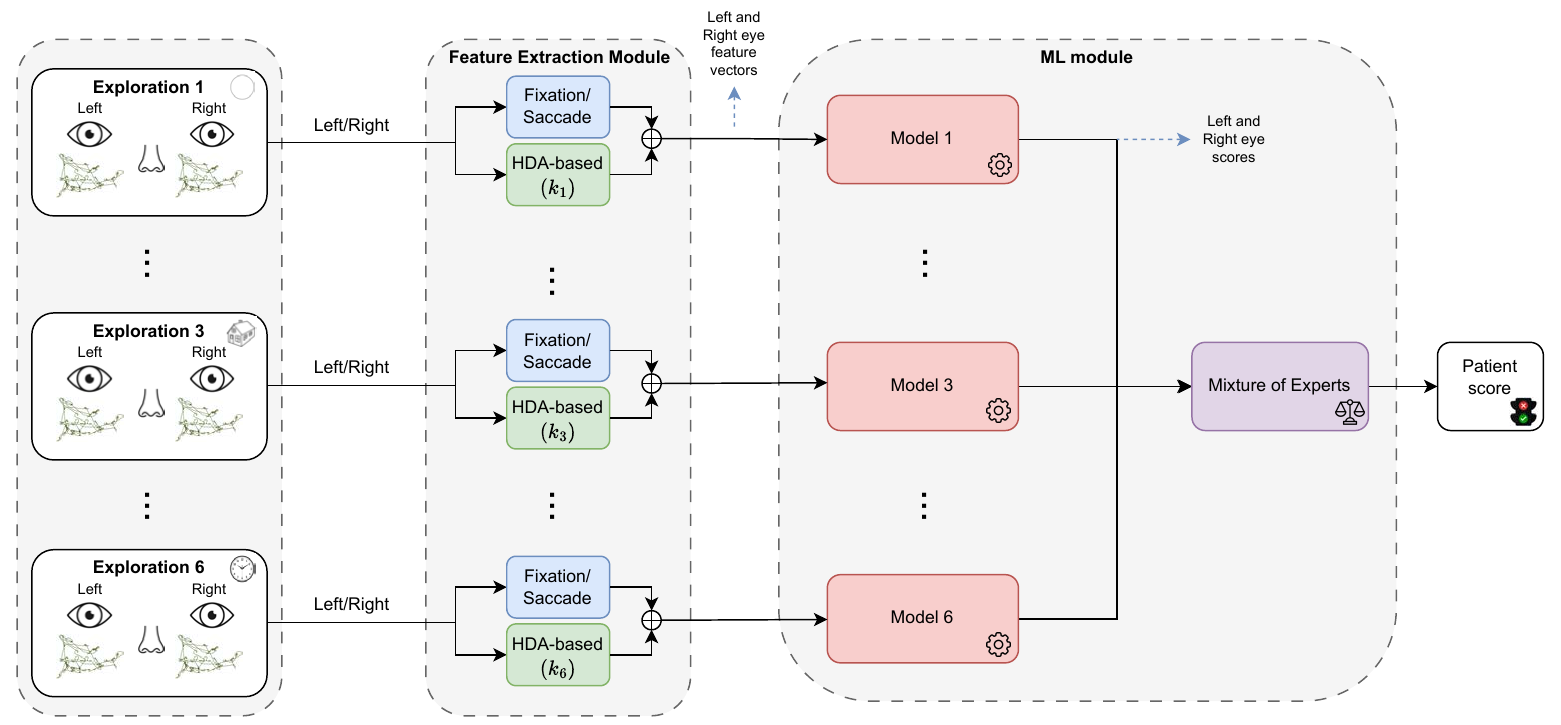}
    \caption{Overview of the classification pipeline. Eye-tracking data is collected from six visual explorations, followed by feature extraction and model training, which yields one score per eye and per exploration. Scores are fused following an \ac{MoE} strategy to yield a final patient-level score.}
    \label{fig:process_pipeline}
\end{figure}

The full pipeline followed is summarised in Fig.~\ref{fig:process_pipeline}, which presents a schematic overview of the overall process from raw gaze recordings through feature extraction, exploration-level classification, and integration of such scores via an ensemble model to obtain a final patient-level score.

\subsection{Feature Extraction: Characterisation of the Visual Exploration Patterns}

This subsection presents the feature extraction strategies used to model visual exploration behaviour from eye-tracking data. Two types of descriptors were computed: fixation/saccade-based metrics, and \ac{HDA}-based features obtained from an unsupervised clustering of gaze data. While the former quantify general oculomotor and spatial behaviours, the latter aim to capture the temporal structure of gaze dynamics via automatic identification of latent \acp{ROI}.

\subsubsection{Fixation/Saccade-Based Features}
\label{subsubsec:fixation_features}

The literature identifies several widely used metrics designed to characterize eye movements during exploration tasks. Among the most common are \ac{TS}, \ac{TSE}, \ac{TF}, \ac{AFT}, \ac{TSA}, and \ac{LD}. These metrics aim to quantify both oculomotor activity and spatial exploration behavior, with particular interest in their potential for automated computation.

The following features are computed independently for each eye and each visual exploration. All metrics are derived from fixation and saccade events automatically identified by the eyetracker device.

\textbf{Total Saccades (TS)} \\ Refers to the number of saccadic eye movements detected during the 15-second exploration window. A saccade is defined as a rapid movement between fixations, delineated by the gaze position at the onset and offset of each fixation, and captures the total count of such transitions. Next is referred to as $TS$.

\textbf{Total Saccadic Excursion (TSE)} \\  
Measures the total distance traversed during all saccadic eye movements detected in the 15-second exploration window. For each of the $TS$ identified saccades, the Euclidean distance between the first and last gaze position is computed. Let each saccade be defined by its initial and final points $\{(x_i^{\text{start}}, y_i^{\text{start}}), (x_i^{\text{end}}, y_i^{\text{end}})\}$ for $i = 1, \dots, TS$, then:
\[
    \mathrm{TSE} = \sum_{i=1}^{TS} \sqrt{(x_i^{\text{end}} - x_i^{\text{start}})^2 + (y_i^{\text{end}} - y_i^{\text{start}})^2}
\]

\textbf{Total Fixations (TF)} \\
Denotes the total number of fixation events detected during the 15-second exploration window. Next is referred to as $TF$. A fixation corresponds to a time period during which the gaze remains relatively stable on a single region of interest.

\textbf{Average Fixation Time (AFT)} \\
Measures the mean duration of all fixations. Assuming that $d_1, d_2, \dots, d_{TF}$ are the durations in milliseconds of the $TF$ fixations during the exploration, then:
\[
    \mathrm{AFT} = \frac{1}{TF} \sum_{m=1}^{TF} d_m
\]

\textbf{Total Scanned Area (TSA)} \\
Captures the spatial extent of exploration by computing the area of the convex hull enclosing all gaze points recorded during the 15-second exploration window. Let $\mathcal{G} = \{(x_1, y_1), (x_2, y_2), \dots, (x_N, y_N)\}$ denote the set of gaze coordinates collected throughout the trial, and let $\mathrm{CH} = \mathrm{Conv}(\mathcal{G})$ be its convex hull. The $TSA$ is defined as the two-dimensional Lebesgue measure of this region:
\[
    \mathrm{TSA} = \lambda(\mathrm{CH})
\]

\textbf{Longest Diagonal (LD)} \\
Measures the maximum straight-line distance between any two gaze points within the convex hull. Let $\mathcal{G} = \{(x_1, y_1), (x_2, y_2), \dots, (x_N, y_N)\}$ denote the set of gaze coordinates, as previously defined. Formally, $LD$ is computed as:
\[
    \mathrm{LD} = \max_{(i,j)} \sqrt{(x_i - x_j)^2 + (y_i - y_j)^2}, \quad i \neq j; \; (x_i, y_i), (x_j, y_j) \in \mathcal{G}
\]
This feature provides a proxy for the furthest visual displacement within the scanned area.

Each feature is typically computed independently for each exploration, participant, and eye. Consequently, the number of fixation/saccade-based features per observation is fixed at 6 per exploration, and is doubled to account for both eyes. 

Table~\ref{tab:features} provides a brief overview of the fixation/saccade features used in this work, along with selected references where similar metrics have been applied.

\begin{table}[]
\caption{Summary of the features used in this work for modelling eye movements during the exploration tasks. The Example Use column includes selected references where similar metrics have been applied.}
\label{tab:features}
\resizebox{\columnwidth}{!}{
\begin{tabular}{ |c|p{4cm}|p{1.5cm}|p{10cm}|  }
\hline
\multicolumn{1}{|l|}{} & \textbf{Parameter}          & \textbf{Ref.} & \textbf{Remarks}                                         \\ \hline
\multirow{6}{*}{{\rotatebox{90}{\textbf{Fixation/Saccade-Based }}}} &
  Total Saccades (TS) &
  \cite{matsumoto2011small} &
  Total number of saccadic events identified during each exploration \\ \cline{2-4} 
 &
  Total Saccadic Excursion (TSE) &
  \cite{matsumoto2011small} &
  Cumulative distance covered during all saccadic events, calculated as the sum of Euclidean distances for all saccades \\ \cline{2-4} 
                       & Total Fixation (TF)         & \cite{matsumoto2011small}            & Total number of fixation events per exploration                                              \\ \cline{2-4} 
                       & Average Fixation Time (AFT) & \cite{matsumoto2011small, wong2020prolonged}            & Mean duration of all fixation events, calculated by averaging the durations of each fixation \\ \cline{2-4} 
 &
  Total Scanned Area (TSA) &
  \cite{matsumoto2011small} &
  Area of the convex hull enclosing all fixation coordinates, representing the overall spatial extent of visual exploration \\ \cline{2-4} 
 &
  Longest Diagonal (LD) &
  \cite{matsumoto2011small, wong2020prolonged} &
  Maximum straight-line distance between any two fixation points within an exploration, used as a proxy for visual spread \\ \hline
\multicolumn{1}{|l|}{\multirow{4}{*}{\rotatebox{90}{\textbf{\ac{HDA}-Based }}}} &
  Fractional Occupancy (FO) &
  \cite{tibon2021transient, bustamante2023classification} &
  The proportion of time spent in each component of the GMM \\ \cline{2-4} 
\multicolumn{1}{|l|}{} & Mean Lifetime (MLT)         & \cite{tibon2021transient, bustamante2023classification}            & The average time that the system stays in each component once it is entered                      \\ \cline{2-4} 
\multicolumn{1}{|l|}{} & Mean Interval Length (MIL)  & \cite{tibon2021transient, bustamante2023classification}           & The average time between recurring visits to each component                                      \\ \cline{2-4} 
\multicolumn{1}{|l|}{} & Entropy of States (EoS)     & \cite{arias2015entropies}           & The Shannon entropy applied to the observed probabilities given the model                    \\ \hline
\end{tabular}%
}
\end{table}

\subsubsection{High-Density Area (HDA)-Based Features}
\label{subsubsec:hda_features}

In this work, a \ac{GMM} is used to model the spatial distribution of gaze positions in a visual exploration. Each gaze position is represented by a two-dimensional point $\mathbf{x}_i=(x_i, y_i) \in \mathbb{R}^2$, corresponding to the horizontal and vertical screen coordinates. The \ac{GMM} defines the probability density function, $p(\mathbf{x}_i)$, of such points as a weighted sum of $k$ Gaussian components:

\begin{equation}
p(\mathbf{x}_i) = \sum_{\kappa=1}^{k} \pi_\kappa \, \mathcal{N}(\mathbf{x}_i \mid \boldsymbol{\mu}_\kappa, \boldsymbol{\Sigma}_\kappa),
\end{equation}

where each component $\kappa$ is parametrised by a mean $\boldsymbol{\mu}_\kappa$, a covariance matrix $\boldsymbol{\Sigma}_\kappa$, and a mixing coefficient $\pi_\kappa$, such that $\sum_\kappa \pi_\kappa = 1$ and $\pi_\kappa \geq 0$. These parameters are estimated using the \ac{EM} algorithm \cite{bishop2006pattern}, which maximises the likelihood of the observed data under the mixture model. 

Compared to simpler clustering approaches such as k-means, a \ac{GMM} offers greater flexibility. It allows clusters to overlap and incorporate full covariance matrices, enabling the modelling of clusters with varying shapes, sizes, and orientations. This flexibility is particularly advantageous when modelling gaze data, where fixation patterns may form elongated or irregularly shaped distributions.

\begin{figure}[!ht]
    \centering
    \includegraphics[width=0.4\linewidth]{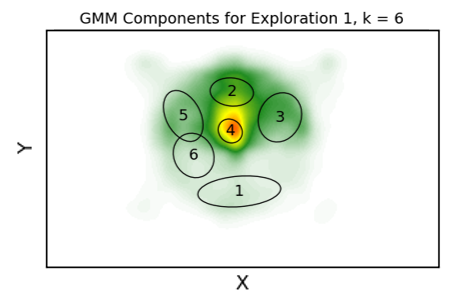}
    \caption{Example of GMM-fitted spatial components for a single exploration task (\textit{Expl.~1}) with $k=6$. The ellipses indicate the learned Gaussian components, while the background shows the empirical density of gaze points. Each component corresponds to an \ac{HDA}, which is identified through unsupervised clustering. Note that the size of the ellipses represents the component covariances, typically showing one standard deviation contour; they do not define hard boundaries, as Gaussian components are nonzero across the entire space. Consequently, all gaze points are probabilistically assigned, and no data is excluded or orphaned.}
    \label{fig:gmm_example}
\end{figure}

Figure~\ref{fig:gmm_example} illustrates a typical example of the fitted Gaussian components for one exploration. Each ellipse represents one of the identified \acp{HDA}, indicating regions where gaze points tend to concentrate. The background heatmap shows the underlying distribution of gaze data, with warmer colours indicating higher point density. This visualisation exemplifies how the \ac{GMM} identifies latent \acp{ROI} directly from the data, without any manual annotation.

Unlike traditional approaches that require manual annotation of regions of interest, the proposed method leverages the gaze data itself to define the most relevant spatial components. While the parameters of each Gaussian component are estimated via the \ac{EM} algorithm, the number of components, $k$, must be specified in advance. However, $k$ is not fixed a priori; instead, it is selected based on the data through an automatic model selection process (see below), enabling generalisation across different images and participants while preserving interpretability. 

Each component is interpreted as an \ac{HDA}, that is, a spatial zone where gaze points are densely concentrated. This terminology differentiates the present model from classical state-based approaches, such as \acp{HMM}, which model temporal transitions and state observations as a joint probability density function. In contrast, the temporal information and \ac{HDA} observations are modelled here as independent processes. Since \acp{HDA} are observable regions with an associated Gaussian density function, the process can be understood as a \ac{MC} with noisy observations where the states are not hidden as in an \ac{HMM}, but blurred. While \acp{HDA} do not constitute formal states, the probabilistic assignment of ordered gaze points to components, according to the components' responsibilities as determined by the \ac{GMM}, enables the construction of temporal state-like sequences, facilitating the extraction of dynamic descriptors of visual exploration. Furthermore, transitions between \acp{HDA} may represent shifts between distinct \acp{ROI} and can offer insight into the underlying dynamics of gaze behaviour. 

Following model fitting, each gaze point is assigned to the component with the highest posterior probability, resulting in a sequence of \ac{HDA} assignments. This sequence is used to compute a set of features, including \ac{FO}, \ac{MLT}, \ac{MIL}, and \ac{EoS}, which have been adapted from other works dedicated to state-based modelling. The formal definitions are presented below.

\textbf{Fractional Occupancy (FO)} \\
Fractional Occupancy measures the proportion of gaze coordinates assigned to a given \ac{HDA} during the 15-second exploration. Let \( N \) be the total number of coordinates, and \( n_\kappa \) the number of samples assigned to component \( \kappa \) via posterior probability. Then:
\[
    \mathrm{FO}_\kappa = \frac{n_\kappa}{N}
\]
This feature reflects the relative dominance of each spatial region during exploration.

\textbf{Mean Lifetime (MLT)} \\
Mean Lifetime represents the average number of consecutive gaze samples (coordinates) assigned to component \( \kappa \) before a transition occurs. Let \( L_\kappa \) be the set of contiguous samples assigned to \( \kappa \), and \( l_m \) the duration (in samples) of each episode. Then:
\[
    \mathrm{MLT}_\kappa = \frac{1}{|L_\kappa|} \sum_{m=1}^{|L_\kappa|} l_m
\]
This metric captures how long attention typically dwells in a given region before moving elsewhere.

\textbf{Mean Interval Length (MIL)} \\
Mean Interval Length quantifies the average number of observations between successive visits to component \( \kappa \). Let \( I_\kappa \) be the set of intervals between visits to \( \kappa \),  and \( d_m \) the duration of each interval. Then:
\[
    \mathrm{MIL}_\kappa = \frac{1}{|I_\kappa|} \sum_{m=1}^{|I_\kappa|} d_m
\]

Higher values suggest that the gaze returns to this region less frequently.

\textbf{Entropy of States (EoS)} \\
Entropy of States computes the Shannon entropy~\cite{shannon1948} of the sequence of \ac{HDA} assignments, providing a global measure of exploration variability. Let \( P_\kappa \) be the empirical proportion of samples assigned to each component \( \kappa \). Then:
\[
    \mathrm{EoS} = -\sum_{\kappa=1}^{k} P_\kappa \log_2 P_\kappa
\]
Higher values of EoS indicate a more diverse and unpredictable allocation of gaze across spatial regions.

Metrics are computed separately for each exploration, patient, eye, and \ac{HDA}; except for EoS, which is estimated for the entire model. In addition, each metric is complemented by two values that describe the maximum and minimum across all \ac{HDA}s. These additional features enable comparisons that are invariant to the arbitrary ordering of components, improving robustness to random initialisation and variation across the cross-validation procedure that will be followed. Consequently, the number of \ac{HDA}-based features per observation is determined by the number of \ac{HDA}. Specifically, for each \ac{HDA}, four features are extracted —\ac{FO}, \ac{MLT}, \ac{MIL}, and \ac{EoS}— for each eye, resulting in $4 \times k \times 2$ features. Additionally, 8 features represent the global maxima and minima of each metric across all states and both eyes. For example, if a single exploration is modelled using 15 \acp{HDA}, the resulting feature set would include $4 \times 15 \times 2 = 120$ \ac{HDA}-based features, plus 8 extrema-based features, yielding a total of 128.

Table~\ref{tab:features} provides a brief overview of the \ac{HDA} features used in this work, along with selected references where similar metrics have been applied. 

\paragraph{Selection of the optimal number of components}

The number of centres (i.e., the number of components of the number of \acp{HDA}) of the \ac{GMM}, $k$, is expected to vary across explorations due to the different characteristics of each image to be explored. Thus, a specific procedure is required to identify this hyperparameter. It is significant to note that a high number of centres present two major risks: a lack of generalisation and the risk of assigning too few data points to each component, making them less meaningful and potentially collapsing the model.

The identification of the most relevant latent regions could be carried out manually. However, to mitigate potential biases and ensure the process is fully automatic, two strategies were developed to determine the number and location of the \acp{HDA}. The first is based on the \ac{BIC}, and the second on an exhaustive search maximising the classification accuracy. 

The first strategy, which employs the \ac{BIC}, uses a penalised likelihood measure that trades off model complexity against data fit \cite{schwarz1978bic}. The optimal number of components, $k$, was selected at the elbow point of the \ac{BIC} curve, beyond which further increases in $k$ yield diminishing returns. For this purpose, the elbow was estimated algorithmically as the point where the slope of the smoothed \ac{BIC} curve approaches \(-1\).

For the second approach, an exhaustive search was conducted to identify the number of components. $k$ was optimised automatically during the cross-validation procedure followed based on the performance of the classifiers. 

\subsection{Machine Learning Classification}
\label{sec:ml_classification}

Several classification algorithms were initially evaluated using lightweight configurations. Two variants of \ac{RF} with 100 trees were used: a shallow version (\ac{RF}-S) with depth 2, and a deeper (\ac{RF}-D) with a maximum depth of 5. Besides, \acp{SVM} were tested with both linear (\ac{SVM}-L) and radial basis function kernels (\ac{SVM}-RBF), using initially the default regularisation parameter ($C = 1.0$) and, for the RBF kernel, the default kernel coefficient ($\gamma = \texttt{"scale"}$). While these parameter choices are not strictly the default values in standard implementations, they were kept fixed across experimental conditions to allow for a fair comparison between model families. A detailed overview of these algorithms can be found in \cite{hastie2009elements}.

The average \ac{AUC} of each classifier type was assessed across all explorations using the cross-validation scheme described in Section~\ref{sec:experimental_setup}. Based on these results, the best-performing classifier was selected for further tuning. A grid search was then carried out to optimise the hyperparameters of the best classifier for each exploration.

\subsection{Combination of Models}
\label{sec:mixture_experts} 


Given that each patient undergoes two eye-based examinations and a separate model is trained for each exploration, discrepancies may occur between the outputs of these models. Therefore, relying solely on individual models is insufficient, and their outputs must be combined and integrated into a unified classifier.

A well-established methodology for combining multiple models is the \ac{MoE} approach, a technique extensively studied in the ensemble learning literature~\cite{sharkey1996combining, breiman1996stacked}. \ac{MoE} frameworks generally outperform single best predictor selection methods~\cite{kittler1998combining}, offering lower variance, and thus lower error, when the ensemble members produce sufficiently diverse outputs to benefit from their integration~\cite{zhou2012ensemble}.

This work focuses on voting-based aggregation strategies, of which two different versions were evaluated. In the unweighted version, all model outputs are averaged equally to produce a final patient-level score. In the weighted version, the contribution of each model is scaled according to its cross-validated performance, typically based on the \ac{AUC}. This approach prioritises more reliable classifiers while still incorporating complementary information from others.

In both cases, the raw probability scores at the exploration-level are used as input to estimate a new patient-level score, which is used to finally carry out the classification. 

\subsection{Exploration Selection}
\label{sec:sec:feature_selection} 

To enhance efficiency and interpretability, a \ac{FFS} \cite{hastie2009elements} is also applied during the model combination phase to determine whether a smaller subset of explorations can match or surpass the classification performance of the entire model. This not only reduces computational load, but also has practical implications, reducing the complexity of the models and/or potentially simplifying the exploration protocol, which decreases both patient burden and physician workload.

\subsection{Experimental setup}
\label{sec:experimental_setup} 

The \ac{HDA}-based features were generated for different values of $k$ ranging from 5 to 50. The objective was to select the optimal number of states that maximises the performance of the classification models. For each $k$, a \ac{GMM} with $k$ components was fitted exclusively on data from the \ac{HC} group, using a fixed random seed to ensure reproducibility. This approach was chosen to define a reference model of healthy visual exploration behaviour, so that deviations in \ac{PD} participants could be interpreted as potential markers of \ac{PD}. At this stage, \ac{yHC} participants were excluded.

After this assignment, the \ac{HDA}-based features were extracted per patient and per exploration. These features capture the spatial-temporal dynamics of gaze in a manner that generalises across image content and does not require manual annotation. In parallel, the set of fixation and saccade-based features was computed; these are independent of $k$ and remain fixed across all experiments. 

Three experimental frameworks (EF) were defined to evaluate the contribution of each feature set. EF1 uses only fixation- and saccade-based metrics. EF2 uses only \ac{HDA}-based features derived from the \ac{GMM} assignments. EF3 combines both feature types.


For those experiments where \(k\) is treated as a tunable hyperparameter, the experiments in EF2 and EF3 are repeated for each value of \(k\), resulting in a distinct feature matrix and a full modelling pipeline for every setting. These are treated as independent experiments rather than as parameter variations of a single model. In contrast, when \(k\) is fixed in advance based on the \ac{BIC} elbow criterion, only a single feature matrix is generated per exploration, and the full modelling pipeline is executed once per experimental framework using that fixed value of \(k\).

The classification task is defined as a binary problem of distinguishing between \ac{PD} and \ac{HC} participants. A 20-fold cross-validation scheme is used to ensure robust and unbiased performance estimation. This cross-validation is also used to jointly select the optimal number of components $k$, the classifier type, and its hyperparameters. An additional configuration was also tested in which the value of $k$ was fixed to the one obtained via BIC, while cross-validation was still used to select the classifier and its parameters. Folds are constructed at the participant level to ensure that all explorations and both eyes from a given individual are assigned entirely to either the training or validation set, thereby avoiding data leakage. Additionally, a separate test set comprising held-out participants is excluded from all training and selection steps and is used solely for final evaluation.

As described in section \ref{sec:methods}, multiple classifiers were initially tested using default parameters. The classifier type with the best average performance across explorations was selected for further tuning of its hyperparameters via grid search. For the \ac{SVM}-RBF, the regularisation parameter \(C\) and the kernel coefficient \(\gamma\) were both varied across 7 values, logarithmically spaced between \(10^{-3}\) and \(10^3\).

To evaluate the ensemble strategies for patient-level scores, a second cross-validation stage is performed using the same 20-folds structure. Here, outputs from individual exploration-level classifiers are aggregated using the \ac{MoE} approaches introduced earlier. Each configuration yields a single probability score per patient. The ensemble strategy (i.e., weighted or unweighted) with the highest average validation performance is retained and evaluated on the held-out test set.

Once the best-performing exploration-level models and the optimal \ac{MoE} strategy for aggregating scores are selected via cross-validation, the complete modelling pipeline is tested on the full training set. The final performance metrics are then computed using the test set held out from the beginning of the process. 

In all cases, decision thresholds for classification were computed on the training folds during cross-validation and were not recalibrated on the test set. Thresholds were selected based on the Equal Error Rate (EER) criterion, defined as the point where false positive and false negative rates are equal.

\section{Results}
\label{sec:results}

\subsection{Descriptive statistics}

A descriptive statistical analysis of the individual features was carried out for the three available cohorts (\ac{yHC}, \ac{HC}, and \ac{PD}). The aim is to provide evidence of the discrimination capabilities of each of the features considered. Full results, including statistical significance tests and visualisations, are presented in Appendix~\ref{sec:appendix}.

The analyses evidenced differences in means for a substantial number of features. These differences are statistically significant between \ac{PD} patients and \ac{HC}, supporting their relevance for cohort discrimination.

\paragraph{Fixation/saccade-based features}

\ac{PD} participants consistently exhibited fewer \ac{TS} than \ac{HC} and \ac{yHC}, with the last showing the highest median values. A consistent decreasing trend was also observed for the \ac{TSE} between cohorts, from \ac{yHC} to \ac{HC} to patients with \ac{PD}. \ac{PD} participants also exhibited fewer \ac{TF} than both \ac{HC} and \ac{yHC}, with the latter showing slightly lower medians than \ac{HC}. Besides, \ac{PD} participants exhibited longer fixation durations than \ac{HC}, with \ac{yHC} also surpassing \ac{HC}. Moreover, the \ac{TSA} decreased progressively from \ac{yHC} to \ac{HC}, and was the lowest for \ac{PD} participants. Similar to \ac{TSA}, the \ac{LD} showed a decreasing pattern from \ac{yHC} to \ac{HC}, and \ac{PD}.

\paragraph{HDA-based features}

Due to the high dimensionality of the \ac{HDA}-based feature space, particularly when the number of \ac{GMM} components, $k$, increases,  the descriptive statistics are limited to a set of illustrative cases. It is also important to note that the component indices of the \ac{GMM} are arbitrarily assigned and vary across initialisations, meaning their numeric labels do not correspond to any fixed spatial or functional order.

For the raw \ac{HDA} features (i.e., \ac{FO}, \ac{MLT}, \ac{MIL}, and \ac{EoS}), fewer consistent trends were observed when examining mean differences directly. Some individual components showed statistical differences, particularly between \ac{PD} and \ac{HC}, but these were not widespread across all explorations. This suggests that raw per-\ac{HDA} metrics may be less reliable when interpreted independently.

In contrast, the summary statistics derived from these features (specifically the maximum and minimum values across \acp{HDA}) showed more robust cohort-level differences. For example, max \ac{FO}, \ac{MIL}, and \ac{EoS} tended to increase from \ac{yHC} to \ac{HC} to \ac{PD}, while min values of all metrics showed a decreasing trend across the same cohorts. These patterns are presented in Appendix~\ref{sec:appendix}.

These results motivate the inclusion of both fixation/saccade-based and \ac{HDA}-based features in the classification framework, as each captures complementary aspects of visual exploration behaviour.

\subsection{Evaluation at exploration level}

This section presents the classification results obtained for each individual exploration. 

The first set of results corresponds to the classifiers introduced in Section~\ref{sec:ml_classification}, evaluated under lightweight configurations, that is, without hyperparameter tuning. The average \acp{AUC} obtained across all explorations and experimental frameworks (EF1, EF2, EF3) are reported in Table~\ref{tab:combined_auc_all}. These values served as a baseline to guide model selection for further optimisation.

\begin{table}[ht]
    \small
    \centering
    \caption{\ac{AUC} Values per Exploration and Experimental Framework}
    \label{tab:combined_auc_all}
    \begin{tabular}{l|l|c|c|c|c}
        \hline
        \textbf{Expl.} & \textbf{EF} & \textbf{RF-S} & \textbf{RF-D} & \textbf{SVM-L} & \textbf{SVM-RBF} \\
        \hline 
\rowcolor{fixsac} \cellcolor{white}  \multirow{3}{*}{}
    & Fix-Sac (EF1) & 0.67 ± 0.11 & 0.64 ± 0.11 & 0.62 ± 0.13 & 0.59 ± 0.12 \\ 
\rowcolor{hda} \cellcolor{white} \textit{Expl.~1} &  \ac{HDA} (EF2)   & 0.62 ± 0.14  & 0.60 ± 0.14  & 0.56 ± 0.15  & 0.56 ± 0.15  \\
\rowcolor{fused} \cellcolor{white} &  Fused (EF3)  & 0.68 ± 0.12  & 0.65 ± 0.13  & 0.65 ± 0.14  & 0.61 ± 0.15  \\
\hline

\rowcolor{fixsac} \cellcolor{white} \multirow{3}{*}{}
     &  Fix-Sac (EF1) & 0.69 ± 0.14 & 0.72 ± 0.13 & 0.63 ± 0.10 & 0.70 ± 0.11 \\
\rowcolor{hda} \cellcolor{white} \textit{Expl.~2}     &  \ac{HDA} (EF2)  & 0.66 ± 0.16  & 0.68 ± 0.16  & 0.64 ± 0.16  & 0.69 ± 0.16  \\
 \rowcolor{fused} \cellcolor{white}    &  Fused (EF3)  & 0.70 ± 0.15  & 0.72 ± 0.15  & 0.65 ± 0.16  & 0.72 ± 0.16  \\
\hline

\rowcolor{fixsac} \cellcolor{white} \multirow{3}{*}{}
     &  Fix-Sac (EF1) & 0.56 ± 0.12 & 0.56 ± 0.17 & 0.49 ± 0.13 & 0.54 ± 0.15 \\
\rowcolor{hda} \cellcolor{white} \textit{Expl.~3}      & \ac{HDA} (EF2)  & 0.54 ± 0.13  & 0.52 ± 0.13  & 0.50 ± 0.13  & 0.50 ± 0.13  \\
\rowcolor{fused} \cellcolor{white}     &  Fused (EF3)  & 0.55 ± 0.12  & 0.54 ± 0.12  & 0.49 ± 0.13  & 0.49 ± 0.13  \\
\hline

\rowcolor{fixsac} \cellcolor{white} \multirow{3}{*}{}
     &  Fix-Sac (EF1) & 0.62 ± 0.11 & 0.65 ± 0.11 & 0.57 ± 0.10 & 0.50 ± 0.11 \\
\rowcolor{hda} \cellcolor{white} \textit{Expl.~4}     &  \ac{HDA} (EF2)  & 0.59 ± 0.13  & 0.56 ± 0.15  & 0.50 ± 0.15  & 0.57 ± 0.14  \\
 \rowcolor{fused} \cellcolor{white}   &  Fused (EF3)  & 0.60 ± 0.12  & 0.58 ± 0.14  & 0.51 ± 0.14  & 0.58 ± 0.13  \\
\hline

\rowcolor{fixsac} \cellcolor{white} \multirow{3}{*}{}
     & Fix-Sac (EF1) & 0.53 ± 0.16 & 0.51 ± 0.15 & 0.42 ± 0.12 & 0.51 ± 0.13 \\
\rowcolor{hda} \cellcolor{white} \textit{Expl.~5}     & \ac{HDA} (EF2)  & 0.55 ± 0.13  & 0.53 ± 0.14  & 0.53 ± 0.14  & 0.56 ± 0.13  \\
 \rowcolor{fused} \cellcolor{white}     & Fused (EF3)  & 0.54 ± 0.12  & 0.53 ± 0.13  & 0.53 ± 0.14  & 0.57 ± 0.12  \\
\hline

\rowcolor{fixsac} \cellcolor{white} \multirow{3}{*}{}
     &  Fix-Sac (EF1) & 0.66 ± 0.15 & 0.64 ± 0.15 & 0.62 ± 0.12 & 0.68 ± 0.11 \\
\rowcolor{hda} \cellcolor{white} \textit{Expl.~6}     &  \ac{HDA} (EF2)  & 0.57 ± 0.16  & 0.56 ± 0.15  & 0.54 ± 0.13  & 0.55 ± 0.14  \\
 \rowcolor{fused} \cellcolor{white}     &  Fused (EF3)  & 0.60 ± 0.17  & 0.59 ± 0.17  & 0.55 ± 0.14  & 0.57 ± 0.15  \\
\hline
    \end{tabular}
\end{table}

These initial results (Table~\ref{tab:combined_auc_all}) suggest differences in performance for EF2 and EF3, which appear to be consistent across models. For example, \textit{Expl.~3} (the house) and \textit{Expl.~5} (the Rey-Osterreith complex figure) consistently show the lowest \ac{AUC}, even falling below 0.5 in some cases. In contrast, \textit{Expl.~2} performs significantly better, with \acp{AUC} often exceeding 0.7.

For EF2 and EF3, the values reported correspond to averages across all tested values of \(k\), providing a general overview of how HDA-based features behave across explorations. While this aggregation does not reflect any specific model configuration, it offers useful insights into the relative discriminative potential of each exploration when incorporating latent spatial descriptors.

In view of these results, and based on the average performance across explorations, the best EF was EF3, and the best classifier was an \ac{SVM}-RBF, so this configuration was selected for further experimentation.

Next, the hyperparameters of the model were adjusted using an exhaustive search following the strategy presented in section \ref{sec:experimental_setup}. The optimal regularisation parameter $C$, number of states $k$, and average \ac{AUC}, are reported in Table~\ref{tab:svm_results} in Appendix~\ref{sec:appendix_hyperparams}. In view of the results and to simplify the model and reduce potential overfitting, the kernel coefficient \(\gamma\) was fixed at 0.01 across all explorations. This choice yielded comparable results to those obtained when optimising \(\gamma\), with only a marginal drop in \ac{AUC} observed in \textit{Expl.~4} (from 0.71 to 0.70), while the average \ac{AUC} for all other explorations remained unchanged.

Once the hyperparameters of the model were adjusted, \textit{Expl.~2} (the cube) showed the best standalone performance, with an average \ac{AUC} of 0.81. \textit{Expl.~1, 4} and \textit{6} had an average \ac{AUC} higher than 0.70, although \textit{Expl.~1} had a slightly lower standard deviation. The lowest-scoring models, \textit{Expl. 3} and \textit{5}, on average, performed at more than 10 absolute points lower than the best-performing model. 

For detailed results of the per-exploration \ac{SVM} and \ac{RF} configurations, see Table~\ref{tab:svm_results} and Table~\ref{tab:bic_rf_results} in Appendix~\ref{sec:appendix_hyperparams}.

Alternatively, a separate set of experiments was conducted to evaluate the utility of the \ac{BIC}-based selection of \(k\), in which the number of components was fixed to the value of the elbow identified for each exploration. 

Figure~\ref{fig:bic_gmm} (Appendix~\ref{sec:appendix_bic}) shows the \ac{BIC} curves and the elbow values estimated for each exploration. 

Using the \ac{BIC} criterion to select the number of components \(k\), \ac{RF} classifiers consistently yielded the best results among all tested models, as shown in Table~\ref{tab:bic_rf_results}. However, the average performance across explorations was lower than that obtained when \(k\) was treated as a tunable hyperparameter via cross-validation. Specifically, the AUC scores for BIC-based models ranged from 0.60 to 0.75 depending on the exploration, whereas the best SVM-RBF models with tuned \(k\) achieved AUCs between 0.61 and 0.81 (see Table~\ref{tab:svm_results}). Given this consistent performance gap, the \ac{BIC} method was not retained in the final pipeline.


\subsection{Evaluation at patient level}
\label{sec:evaluation_patient_level}  

This section presents the results at the patient level, fusing the information provided by all explorations. 

The raw probability scores from the exploration-level \ac{SVM}-RBF classifiers (which range between 0 and 1) were used as input for the \ac{MoE} voting strategies, akin to a training matrix for both modelling pipelines. Following best practices in medical decision making, tie-breaking is resolved in favour of the positive class (i.e., classifying as \ac{PD}), given that false negatives are costlier than false positives in clinical contexts. 

Table~\ref{tab:moe_best} shows the best-performing \ac{MoE} configuration, which applies an \ac{AUC}-weighted voting strategy across selected exploration-eye outputs. These outputs were selected using a \ac{FFS} strategy, which iteratively added exploration-eye classifiers to the ensemble based on their cross-validated contribution to patient-level \ac{AUC}. Specifically, the final ensemble integrates scores from \textit{Expl.}~1, 2, 5, and 6. This ensemble achieves an average \ac{AUC} of \(0.87 \pm 0.11\), outperforming any individual exploration classifier.

\begin{table}[h!]
    \centering
    \caption{Exploration-eye combinations selected in the final \ac{MoE} configuration. The table indicates which exploration tasks and eye recordings (right or left) contributed to the final ensemble. Only a subset of all available exploration-eye classifiers was retained, based on their individual discriminative performance.}
    \label{tab:moe_best}
    \begin{tabular}{lcc}
        \toprule
        \textbf{Exploration} & \textbf{Right Eye} & \textbf{Left Eye} \\
        \midrule
        \textit{Expl.~1} &  & \checkmark \\
        \textit{Expl.~2} &  & \checkmark \\
        \textit{Expl.~3} &  &  \\
        \textit{Expl.~4} & &  \\
        \textit{Expl.~5} & \checkmark &  \\
        \textit{Expl.~6} & & \checkmark \\
        \bottomrule
    \end{tabular}
\end{table}

\begin{figure} 
    \centering
    \includegraphics[width=0.6\textwidth]{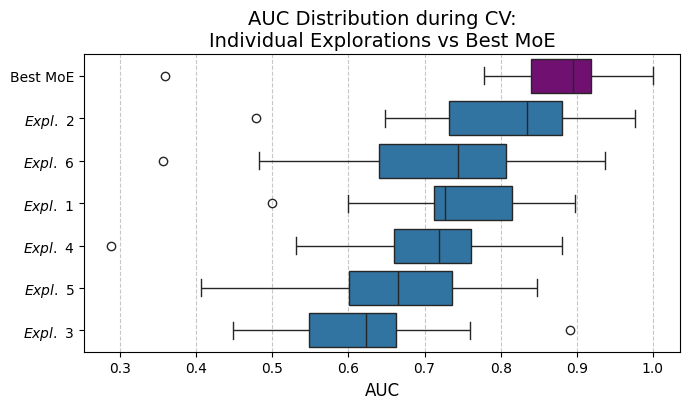}
    \caption{Comparison of the cross-validated \ac{AUC} obtained for the best individual models and \ac{MoE}. The \ac{MoE} strategy shows improved performance and reduced variance.}
    \label{fig:boxplots_individual_vs_moe}
\end{figure}

The \ac{MoE} strategy, based on score-level voting, unifies exploration-level outputs into a single patient-level decision. As shown in Fig.~\ref{fig:boxplots_individual_vs_moe}, the final configuration using AUC-weighted voting achieved both higher and more stable performance across cross-validation folds than any individual exploration model. Specifically, the MoE attained an average \ac{AUC} of \(0.87 \pm 0.11\), compared to \(0.81 \pm 0.12\) for \textit{Expl.~2}, which was the best standalone classifier.

\subsection{Validation of results}
\label{sec:validation}


The \ac{FFS} feature selection procedure consistently identified \textit{Expl.}~1, 2, 5, and 6 as informative. Thus, the model corresponding to \textit{Expl.}~3 and 4 were excluded from the final patient-level score. This suggests that certain images may not contribute additional discriminative information and raises the possibility of simplifying the exploration protocol by discarding tasks with lower predictive value.

The final model, evaluated on the held-out test set, achieved an \ac{AUC} of 0.93, an F1 score of 0.71, a sensitivity of 0.56, and a specificity of 1.00.

\begin{figure}[ht]
    \centering
    \includegraphics[width=0.5\textwidth]{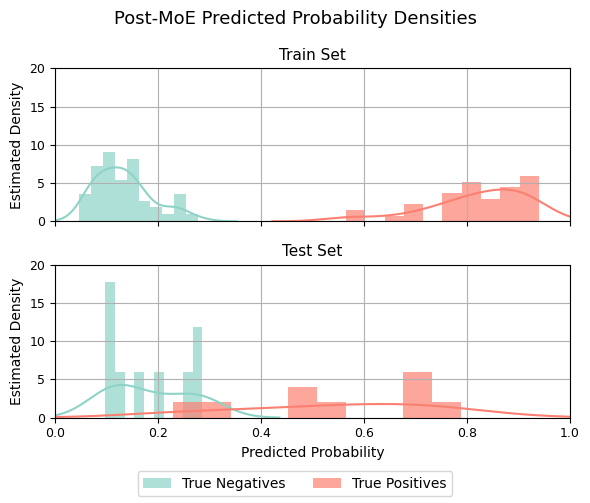}
    \caption{Distributions of final scores (post-\ac{MoE}) for true positive and true negative cases in the training and test sets. In the test set, class separation is less distinct, indicating reduced model confidence and a potential distribution shift.}
    \label{fig:moe_score_distributions}
\end{figure}

To investigate this further, the distribution of the final scores was analised for both true positives and true negatives between the training and test sets. Fig.~\ref{fig:moe_score_distributions} shows the density distributions for each class.

In the training set, positive and negative scores are well separated, with limited overlap between classes. In contrast, in the test set, the class distributions drift closer together, and the separation margin narrows considerably. This shift indicates a potential domain misalignment, likely arising from a certain overfitting to specific characteristics present in the training data. 

\section{Discussion and Conclusions}
\label{sec:conclussion}

This work presents a novel approach for the screening of \ac{PD} using eye tracking data collected during visual exploration of structured images. The pipeline combines classical oculomotor metrics with \ac{HDA}-based features extracted through gaze clustering and, critically, applies machine learning classification models to assess their discriminative power. This integration of computational gaze analysis with advanced modelling constitutes a key contribution of the present study.

A central methodological decision in the proposed framework involves the number of components used to model the spatial distribution of gaze via \acp{GMM}. The value of $k$ (which defines the number of \acp{HDA} or implicit \acp{ROI}) shapes the capacity of the model to capture non-static exploration patterns. While classical approaches might rely on human-defined \acp{ROI}, often limited in number and potentially biased by preconceptions about the image structure, our method allows these spatial regions to emerge directly from the data. Although standard techniques such as \ac{BIC} were initially considered, they yielded suboptimal classification performance. Instead, $k$ was treated as a tunable hyperparameter, selected via cross-validation to maximise discriminative power. This data-driven procedure provided more consistent and task-relevant selections of $k$,  enhancing adaptability across different datasets.

The \ac{MoE} framework outperformed individual exploration-level models, achieving an \ac{AUC} of 0.95 on a held-out test set. Notably, exploration-specific performance varied considerably, suggesting that some images elicit more discriminative gaze patterns than others. In particular, explorations involving familiar or structurally simple images with salient focal points (e.g., clocks, cubes, intersecting shapes) showed stronger discriminative value. This supports the hypothesis that not all visual tasks are equally informative, and that long, heterogeneous tests may be unnecessary. The automatic feature selection process, which consistently favoured a stable subset of explorations, further reinforces this idea. A qualitative consideration of the selected explorations suggests a potential link between the image content and its predictive power. Specifically, images with a familiar structure or an obvious focal point, such as the corner of a cube (\textit{Expl.~2}) or the hands of a clock (\textit{Expl.~6}), may guide gaze behaviour more consistently among participants. This may explain their repeated selection during the feature selection process.

These results suggest that the model is relatively cautious in screening for \ac{PD}, showing stronger performance in identifying negative cases. While this yields high specificity, it also reflects a trade-off that may lead to missed true positives — particularly in early or subtle stages of the disease. In clinical contexts, such conservative thresholds may need to be adjusted depending on application and risk tolerance.

Prior work has shown that visual saliency and semantic familiarity can influence attention and reduce variability in scanpaths~\cite{borji2013state}. This aligns with the present finding that structurally simple or familiar images may elicit more stereotyped gaze patterns. For instance, the hands of a clock may act as culturally and functionally meaningful cues, potentially constraining attention and reducing scanpath variability. Conversely, abstract images with no salient focal points may induce more diffuse and harder-to-model gaze behaviour. From an applied perspective, this raises the possibility of streamlining the protocol by identifying and removing images that do not contribute with an incremental value to the final prediction. In this study, \textit{Expl.}~3 was consistently excluded from the final patient-level score, suggesting that it may be a candidate for removal. However, additional validation with larger datasets will be required to confirm whether such simplifications are justified without compromising model performance or generalisability. Leveraging ensemble strategies such as \ac{MoE} further contributes to a more robust and stable patient-level decision process, underscoring the value of combining outputs across multiple well-chosen visual tasks.

Several deliberate strategies were implemented to mitigate overfitting and enhance reproducibility. These included patient-level cross-validation, strict separation of the test set from all stages of model development, and the use of low-parameter ensemble aggregation (e.g., weighted voting). Additionally, the pipeline avoided manual annotation of regions of interest by relying on probabilistic clustering to define \acp{HDA}. These design decisions aimed to reduce human bias and ensure that performance estimates reflected real-world conditions. Nonetheless, the presence of a distribution shift between training and test sets, evidenced by a drop in sensitivity and narrower class separation, highlights the need for continued refinement. Although the model achieves strong discrimination on held-out data, its generalisation capacity remains inconclusive, particularly in settings that require high-confidence decision thresholds.

Although the current model shows potential, limitations remain in terms of generalisation and sensitivity. Future work should explore alternative strategies for modelling the \ac{ROI}, including more flexible approaches such as neural networks or graph-based architectures that capture spatial dependencies more explicitly. In addition, combining gaze data from both eyes into a single cyclopean eye representation may offer a more stable and perceptually grounded signal than treating each eye separately \cite{luque2024estimation}. Finally, validating the framework on demographically or culturally distinct populations may reveal important variations in visual exploration behaviour and improve model robustness.

Although discriminative power is clearly present, additional work is needed to reach a level of performance that would justify clinical application. Overall, these findings support the feasibility of using noninvasive, short-duration visual exploration tasks as a digital biomarker for neurodegenerative disease and lay the foundation for further development of interpretable, gaze-driven screening tools for \ac{PD}.

\section{Acknowledgments}

This work was supported by the Ministry of Economy and Competitiveness of Spain under Grants PID2021-128469OB-I00 and TED2021-131688B-I00, and by Comunidad de Madrid, Spain. Universidad Politécnica de Madrid also supports Julián D. Arias-Londoño through a María Zambrano UP2021-035 grant funded by European Union-NextGenerationEU. Finally, the authors would like to thank the Madrid ELLIS unit (European Laboratory for Learning \& Intelligent Systems) for its indirect support, and all patients who selflessly participated in the study.

\subsection*{Ethics declaration} 

The study was approved by the Ethics Review Board of the \ac{HUF} and \ac{HGUGM} with codes 18/11-ENM1 and 11/2015 respectively, and in accordance with the Spanish Ethical Review Act. All patients and controls followed the same experimental protocol.

All participants were informed about the project objectives and, if they agreed with the study conditions, they were recruited for participation and recording at \ac{HGUGM} facilities. Participants received a document containing details about the project goals prior to recording. Subsequently, they were asked to sign a consent form. Participants did not receive any compensation for participating in the study, agreeing to share their voices for research purposes. Patients were informed of their rights and the option to leave the study at any time.

Patients were individually identified with a code, which is different from the one used in the Hospital for their clinical histories. No personal data was exchanged with external researchers who had access to the corpus. Only one specialist got in contact with each patient, being also in charge of collecting the clinical data.

\section*{CRediT Author Statement}

J.D.A. and J.I.G. proposed the methodology. M.A.D., J.D.A. and J.I.G. designed the experiment. M.A.D. developed the software to analyse the data. M.A.D., J.D.A. and J.I.G. validated the experiment. J.I.G. provided the resources. M.A.D. wrote the initial draft version. J.D.A. and J.I.G. reviewed and edited the manuscript. J.D.A. and J.I.G. supervised. J.I.G. acquired the funding. All authors have read and agreed to the published version of the manuscript.

\section*{Declaration of Competing interests} 

The authors declare that they have no competing interests.

\newpage
\appendix
\section{Appendix}
\label{sec:appendix}

\subsection{Descriptive statistical analysis}

This section complements the main results by presenting detailed statistical comparisons and visual summaries of the features used for cohort discrimination.

T-tests were performed across three pairwise comparisons: \ac{PD} vs. \ac{HC}, \ac{PD} vs. all controls, and \ac{HC} vs. \ac{yHC}. Statistically significant results ($p < 0.05$) are highlighted in bold in the following tables.

\subsubsection{Fixation/saccade-based metrics}

\begin{figure}
    \centering
    \includegraphics[width=0.9\textwidth]{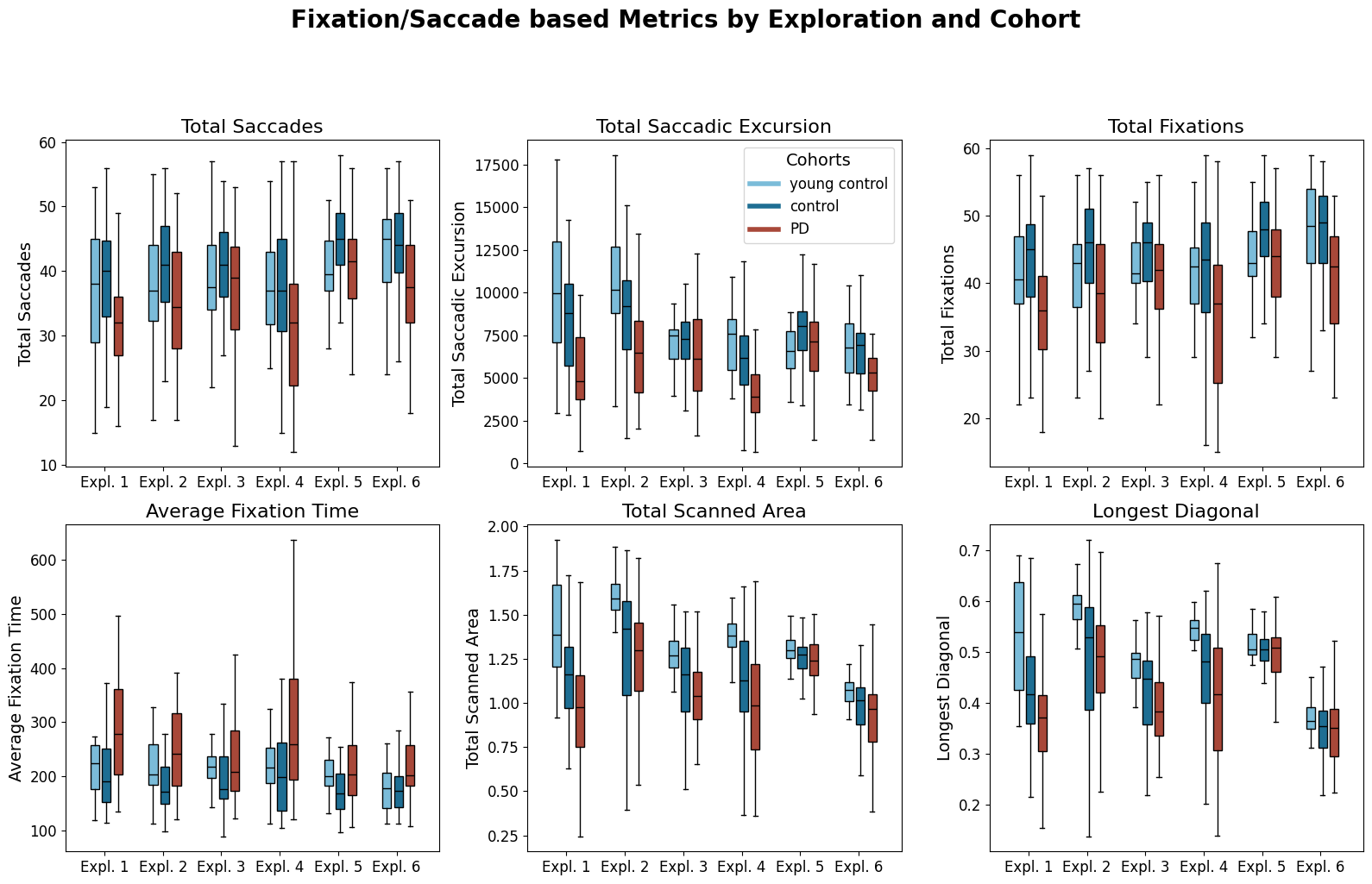}
    \caption{Boxplots for Fixation and Saccadic Metrics across cohorts for the different explorations (Expl. 1-6).}
    \label{fig:fixation_saccade_boxplots}
\end{figure}

Figure~\ref{fig:fixation_saccade_boxplots} presents boxplots of fixation/saccade features for all six explorations and across the three cohorts. Notably, \textit{Expl.~4} (intersecting pentagons) exhibited atypical trends and was excluded from statistical comparisons discussed in the following.

Differences in means were most pronounced between \ac{PD} and \ac{HC}, and between \ac{PD} and all controls. However, significant differences were also found between \ac{HC} and \ac{yHC}, especially for saccade-based features. This aligns with existing literature on ageing and visual behaviour~\cite{dowiasch2015aging}.

When testing the differences in means by combining all controls (i.e., \ac{yHC} and \ac{HC}) with \ac{PD} patients, most features remain statistically significant. This reinforces the notion that these variables might possess discriminatory power to differentiate between \ac{HC} and \ac{PD} cohorts. 

Table~\ref{tab:pvals_pd_hc_allctrl} summarises the statistical significance for the fixation/saccade-based features across all explorations.

\begin{table}[ht]
    \centering
    \scriptsize
    \caption{T-test p-values for fix/sac features across six visual explorations for each cohort}
    \label{tab:pvals_pd_hc_allctrl}
    \begin{tabular}{|c|c|c|c|c|c|c|c|}
    \hline
    \textbf{Exploration} & \textbf{Compared Cohorts} & \textbf{\makecell{Average\\Fixation Time}} & \textbf{Longest Diagonal} & \textbf{Total Fixations} & \textbf{Total Saccades} & \textbf{\makecell{Total\\Saccadic Excursion}} & \textbf{\makecell{Total\\Scanned Area}} \\
    \hline

\multirow{3}{*}{\textit{Expl.~1}}
    & PD vs HC       & \textbf{2.33 $\times10^{-4}$} & \textbf{4.28 $\times10^{-3}$} & \textbf{3.34 $\times10^{-8}$} & \textbf{6.99 $\times10^{-8}$} & \textbf{2.61 $\times10^{-7}$} & \textbf{1.98 $\times10^{-3}$} \\
    & yHC vs HC      & 2.00 $\times10^{-1}$ & \textbf{4.28 $\times10^{-3}$} & 1.08 $\times10^{-1}$ & 1.80 $\times10^{-1}$ & \textbf{2.04 $\times10^{-2}$} & \textbf{1.74 $\times10^{-7}$} \\
    & PD vs All Ctrl & \textbf{4.44 $\times10^{-4}$} & \textbf{8.21 $\times10^{-7}$} & \textbf{1.04 $\times10^{-7}$} & \textbf{2.12 $\times10^{-7}$} & \textbf{2.57 $\times10^{-10}$} & \textbf{7.26 $\times10^{-8}$} \\
\hline

\multirow{3}{*}{\textit{Expl.~2}}
    & PD vs HC       & \textbf{3.93 $\times10^{-7}$} & 9.57 $\times10^{-1}$ & \textbf{2.55 $\times10^{-7}$} & \textbf{1.67 $\times10^{-5}$} & \textbf{3.37 $\times10^{-4}$} & 7.60 $\times10^{-1}$ \\
    & yHC vs HC      & \textbf{1.01 $\times10^{-2}$} & \textbf{4.30 $\times10^{-9}$} & \textbf{4.34 $\times10^{-3}$} & \textbf{4.09 $\times10^{-2}$} & \textbf{1.21 $\times10^{-2}$} & \textbf{1.14 $\times10^{-10}$} \\
    & PD vs All Ctrl & \textbf{3.10 $\times10^{-5}$} & 7.55 $\times10^{-2}$ & \textbf{1.20 $\times10^{-5}$} & \textbf{1.41 $\times10^{-4}$} & \textbf{3.00 $\times10^{-6}$} & \textbf{1.18 $\times10^{-2}$} \\
\hline

\multirow{3}{*}{\textit{Expl.~3}}
    & PD vs HC       & 1.72 $\times10^{-1}$ & 5.35 $\times10^{-2}$ & \textbf{8.38 $\times10^{-3}$} & \textbf{2.18 $\times10^{-2}$} & \textbf{2.66 $\times10^{-2}$} & 8.16 $\times10^{-2}$ \\
    & yHC vs HC      & 5.53 $\times10^{-1}$ & \textbf{1.62 $\times10^{-6}$} & 2.14 $\times10^{-1}$ & 5.14 $\times10^{-1}$ & 5.82 $\times10^{-1}$ & \textbf{3.11 $\times10^{-7}$} \\
    & PD vs All Ctrl & 1.56 $\times10^{-1}$ & \textbf{3.43 $\times10^{-4}$} & \textbf{1.50 $\times10^{-2}$} & \textbf{2.43 $\times10^{-2}$} & \textbf{2.44 $\times10^{-2}$} & \textbf{2.90 $\times10^{-4}$} \\
\hline

\multirow{3}{*}{\textit{Expl.~4}}
    & PD vs HC       & \textbf{3.22 $\times10^{-4}$} & \textbf{2.35 $\times10^{-2}$} & \textbf{1.00 $\times10^{-5}$} & \textbf{1.19 $\times10^{-4}$} & \textbf{7.00 $\times10^{-6}$} & \textbf{1.23 $\times10^{-2}$} \\
    & yHC vs HC      & 6.47 $\times10^{-1}$ & \textbf{3.15 $\times10^{-8}$} & 5.76 $\times10^{-1}$ & 9.28 $\times10^{-1}$ & \textbf{1.39 $\times10^{-3}$} & \textbf{8.51 $\times10^{-9}$} \\
    & PD vs All Ctrl & \textbf{5.76 $\times10^{-5}$} & \textbf{3.14 $\times10^{-3}$} & \textbf{3.43 $\times10^{-6}$} & \textbf{2.23 $\times10^{-5}$} & \textbf{5.44 $\times10^{-9}$} & \textbf{1.63 $\times10^{-5}$} \\
\hline

\multirow{3}{*}{\textit{Expl.~5}}
    & PD vs HC       & \textbf{1.87 $\times10^{-2}$} & 9.80 $\times10^{-1}$ & \textbf{3.04 $\times10^{-4}$} & \textbf{1.08 $\times10^{-3}$} & \textbf{1.88 $\times10^{-2}$} & 7.51 $\times10^{-1}$ \\
    & yHC vs HC      & 1.00 $\times10^{-1}$ & \textbf{3.73 $\times10^{-3}$} & \textbf{8.53 $\times10^{-3}$} & \textbf{2.62 $\times10^{-3}$} & \textbf{5.50 $\times10^{-5}$} & \textbf{4.81 $\times10^{-4}$} \\
    & PD vs All Ctrl & \textbf{2.89 $\times10^{-2}$} & 3.67 $\times10^{-1}$ & \textbf{2.34 $\times10^{-3}$} & \textbf{1.44 $\times10^{-2}$} & 1.73 $\times10^{-1}$ & 1.67 $\times10^{-1}$ \\
\hline

\multirow{3}{*}{\textit{Expl.~6}}
    & PD vs HC       & \textbf{3.58 $\times10^{-2}$} & 5.24 $\times10^{-1}$ & \textbf{5.00 $\times10^{-6}$} & \textbf{4.10 $\times10^{-5}$} & \textbf{4.33 $\times10^{-4}$} & 2.06 $\times10^{-1}$ \\
    & yHC vs HC      & 4.78 $\times10^{-1}$ & \textbf{4.15 $\times10^{-2}$} & 7.09 $\times10^{-1}$ & 9.24 $\times10^{-1}$ & 6.61 $\times10^{-1}$ & \textbf{7.01 $\times10^{-4}$} \\
    & PD vs All Ctrl & \textbf{5.05 $\times10^{-3}$} & 1.51 $\times10^{-1}$ & \textbf{1.57 $\times10^{-7}$} & \textbf{4.75 $\times10^{-6}$} & \textbf{8.90 $\times10^{-5}$} & \textbf{1.16 $\times10^{-2}$} \\
\hline

    \end{tabular}
\end{table}

\subsubsection{HDA-based metrics}

Each combination of exploration and number of components $k$ produces a set of $(k \times 4) + 8$ features, resulting in high dimensionality. As such, detailed analysis is provided using \textit{Expl.~1} as a representative case, since its optimal $k$ is low and suitable for visualisation and statistical comparisons.

It is worth noting that the labels assigned to the $k$ different components are arbitrary and dependent on random initialisation; therefore, their absolute numbering carries no semantic meaning or order.

\begin{figure} 
    \centering
    \includegraphics[width=0.9\textwidth]{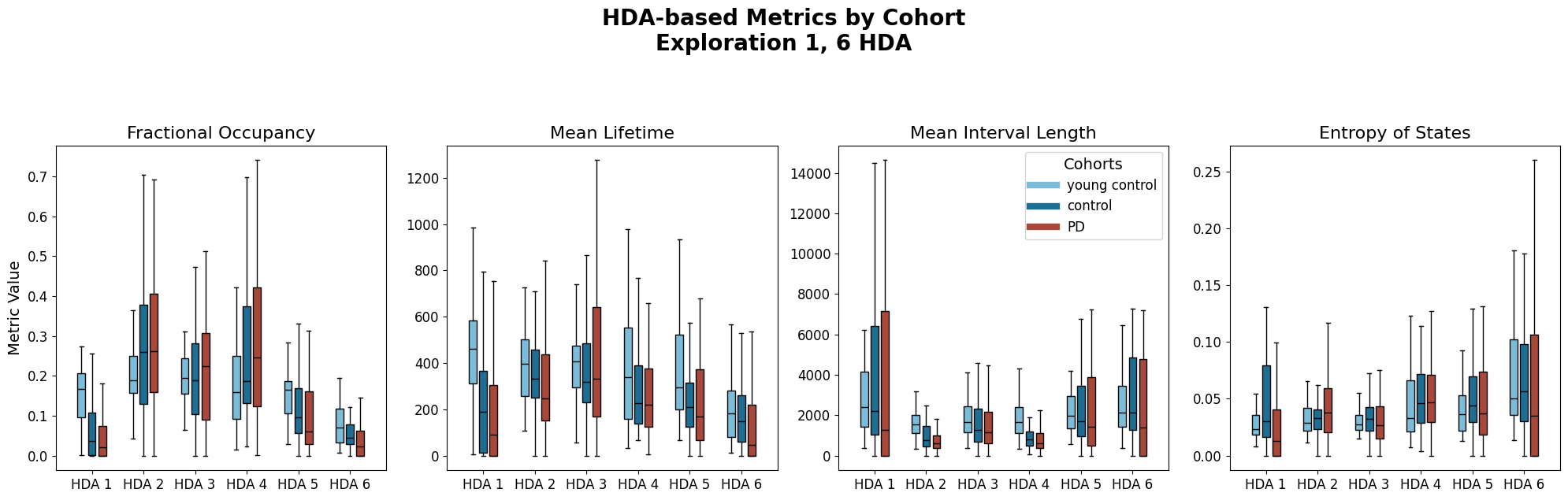}
    \caption{Boxplots for HDA-based Metrics for Expl. 1, $k=6$}
    \label{fig:boxplots_hda_rawmetrics}
\end{figure}

\begin{figure} 
    \centering
    \includegraphics[width=0.9\textwidth]{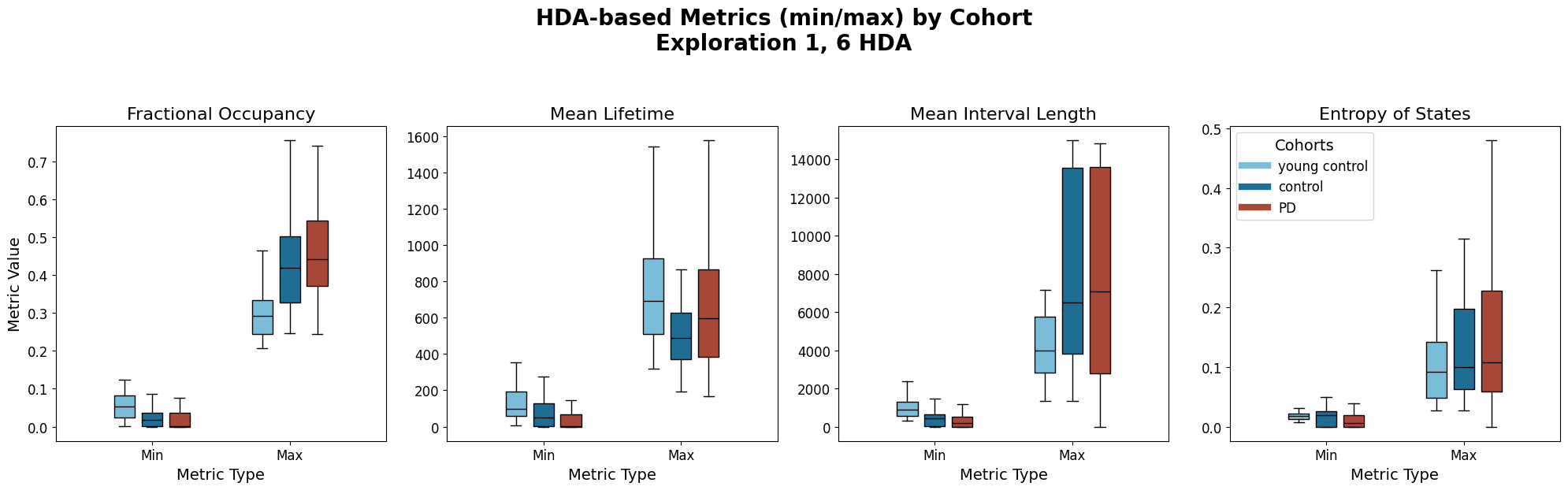}
    \caption{Boxplots for HDA-based Metrics (min/max) for Expl. 1, $k=6$}
    \label{fig:boxplots_hda_minmaxmetrics}
\end{figure}

Figure~\ref{fig:boxplots_hda_rawmetrics} displays the distribution of the raw \ac{HDA} metrics across cohorts for each \ac{HDA}. Figure~\ref{fig:boxplots_hda_minmaxmetrics} aggregates the min and max values of those metrics per exploration, which were found to be more stable and informative.

Table~\ref{tab:pvals_hda} shows the statistical comparisons of the raw \ac{HDA} metrics per component, while Table~\ref{tab:pvals_hda_minmax} presents the same analysis for the aggregated min/max metrics.

Although direct trends across explorations are not meaningful due to the non-identifiability of \ac{HDA}s, within-exploration patterns suggest that several of these features hold discriminatory potential.

\begin{table}[ht]
    \centering
    \scriptsize
    \caption{T-test p-values across for \ac{HDA}-based features \ac{HDA}s for each cohort, Expl. 1, 6 \ac{HDA}s}
    \label{tab:pvals_hda}
    \begin{tabular}{|c|c|c|c|c|c|}
    \hline
    \textbf{\ac{HDA}} & \textbf{Compared Cohorts} & \textbf{\ac{FO}} & \textbf{\ac{MIL}} & \textbf{\ac{MLT}} & \textbf{\ac{EoS}} \\
    \hline

\multirow{3}{*}{\textit{\ac{HDA} 1}} 
    & PD vs HC       & \textbf{1.50$\times10^{-2}$} & 6.74$\times10^{-1}$ & 2.68$\times10^{-1}$ & \textbf{1.12$\times10^{-2}$} \\
    & yHC vs HC      & \textbf{5.78$\times10^{-8}$} & 1.63$\times10^{-1}$ & \textbf{7.99$\times10^{-6}$} & 5.10$\times10^{-2}$ \\
    & PD vs All Ctrl & \textbf{5.29$\times10^{-8}$} & 9.92$\times10^{-1}$ & 5.51$\times10^{-1}$ & \textbf{2.17$\times10^{-2}$} \\
\hline

\multirow{3}{*}{\textit{\ac{HDA} 2}} 
    & PD vs HC       & 2.29$\times10^{-1}$ & \textbf{8.52$\times10^{-3}$} & 9.03$\times10^{-1}$ & 2.59$\times10^{-1}$ \\
    & yHC vs HC      & \textbf{3.77$\times10^{-3}$} & 8.81$\times10^{-1}$ & 2.18$\times10^{-1}$ & 3.25$\times10^{-1}$ \\
    & PD vs All Ctrl & \textbf{2.60$\times10^{-2}$} & \textbf{3.06$\times10^{-4}$} & 6.06$\times10^{-1}$ & 1.09$\times10^{-1}$ \\
\hline

\multirow{3}{*}{\textit{\ac{HDA} 3}} 
    & PD vs HC       & 4.59$\times10^{-1}$ & 7.10$\times10^{-1}$ & \textbf{1.16$\times10^{-1}$} & 1.64$\times10^{-1}$ \\
    & yHC vs HC      & 5.15$\times10^{-1}$ & 3.67$\times10^{-1}$ & 7.93$\times10^{-2}$ & 7.44$\times10^{-1}$ \\
    & PD vs All Ctrl & 3.15$\times10^{-1}$ & 5.14$\times10^{-1}$ & 2.66$\times10^{-1}$ & 1.54$\times10^{-1}$ \\
\hline

\multirow{3}{*}{\textit{\ac{HDA} 4}} 
    & PD vs HC       & 5.41$\times10^{-1}$ & \textbf{6.93$\times10^{-2}$} & 4.58$\times10^{-1}$ & \textbf{1.25$\times10^{-1}$} \\
    & yHC vs HC      & 9.19$\times10^{-2}$ & 2.29$\times10^{-1}$ & 2.57$\times10^{-1}$ & 9.16$\times10^{-1}$ \\
    & PD vs All Ctrl & 1.58$\times10^{-1}$ & \textbf{2.79$\times10^{-3}$} & 5.52$\times10^{-1}$ & 1.35$\times10^{-1}$ \\
\hline

\multirow{3}{*}{\textit{\ac{HDA} 5}} 
    & PD vs HC       & 3.26$\times10^{-1}$ & 7.41$\times10^{-1}$ & 8.21$\times10^{-1}$ & 5.89$\times10^{-1}$ \\
    & yHC vs HC      & \textbf{1.19$\times10^{-2}$} & 1.74$\times10^{-1}$ & \textbf{2.87$\times10^{-4}$} & \textbf{1.04$\times10^{-2}$} \\
    & PD vs All Ctrl & \textbf{4.00$\times10^{-2}$} & 4.55$\times10^{-1}$ & 1.35$\times10^{-1}$ & 8.52$\times10^{-1}$ \\
\hline

\multirow{3}{*}{\textit{\ac{HDA} 6}} 
    & PD vs HC       & \textbf{4.83$\times10^{-3}$} & 9.12$\times10^{-1}$ & 1.34$\times10^{-1}$ & 3.89$\times10^{-1}$ \\
    & yHC vs HC      & 3.98$\times10^{-1}$ & 2.82$\times10^{-1}$ & 3.29$\times10^{-1}$ & 5.79$\times10^{-1}$ \\
    & PD vs All Ctrl & \textbf{1.49$\times10^{-4}$} & 6.33$\times10^{-1}$ & \textbf{4.46$\times10^{-2}$} & 2.87$\times10^{-1}$ \\
\hline

    \end{tabular}
\end{table}

\begin{table}[ht]
    \centering
    \scriptsize
    \caption{T-test p-values for max and min features across groups and \ac{HDA}s. Expl. 1, 6 \ac{HDA}s}
    \label{tab:pvals_hda_minmax}
    \begin{tabular}{|c|c|c|c|c|c|}
    \hline
    \textbf{Aggregation} & \textbf{Compared Cohorts} & \textbf{FO} & \textbf{MIL} & \textbf{MLT} & \textbf{Entropy} \\
    \hline

\multirow{3}{*}{\textit{Max}} 
    & PD vs HC       & 1.25$\times10^{-1}$ & 7.18$\times10^{-1}$ & 6.62$\times10^{-2}$ & 3.04$\times10^{-1}$ \\
    & yHC vs HC      & \textbf{5.17$\times10^{-5}$} & \textbf{1.88$\times10^{-5}$} & \textbf{7.83$\times10^{-3}$} & 6.63$\times10^{-2}$ \\
    & PD vs All Ctrl & \textbf{1.07$\times10^{-3}$} & \textbf{5.48$\times10^{-2}$} & 1.26$\times10^{-1}$ & 9.87$\times10^{-2}$ \\
\hline

\multirow{3}{*}{\textit{Min}} 
    & PD vs HC       & 9.55$\times10^{-2}$ & \textbf{1.09$\times10^{-2}$} & \textbf{4.88$\times10^{-4}$} & \textbf{1.30$\times10^{-2}$} \\
    & yHC vs HC      & \textbf{6.09$\times10^{-6}$} & \textbf{3.55$\times10^{-7}$} & \textbf{2.21$\times10^{-3}$} & 7.98$\times10^{-1}$ \\
    & PD vs All Ctrl & \textbf{2.56$\times10^{-5}$} & \textbf{1.30$\times10^{-7}$} & \textbf{3.04$\times10^{-8}$} & \textbf{3.61$\times10^{-3}$} \\
\hline

    \end{tabular}
\end{table}

\subsection{BIC Curves for Selection of $k$}
\label{sec:appendix_bic}

\begin{figure}[h!]
    \centering
    \includegraphics[width=0.95\textwidth]{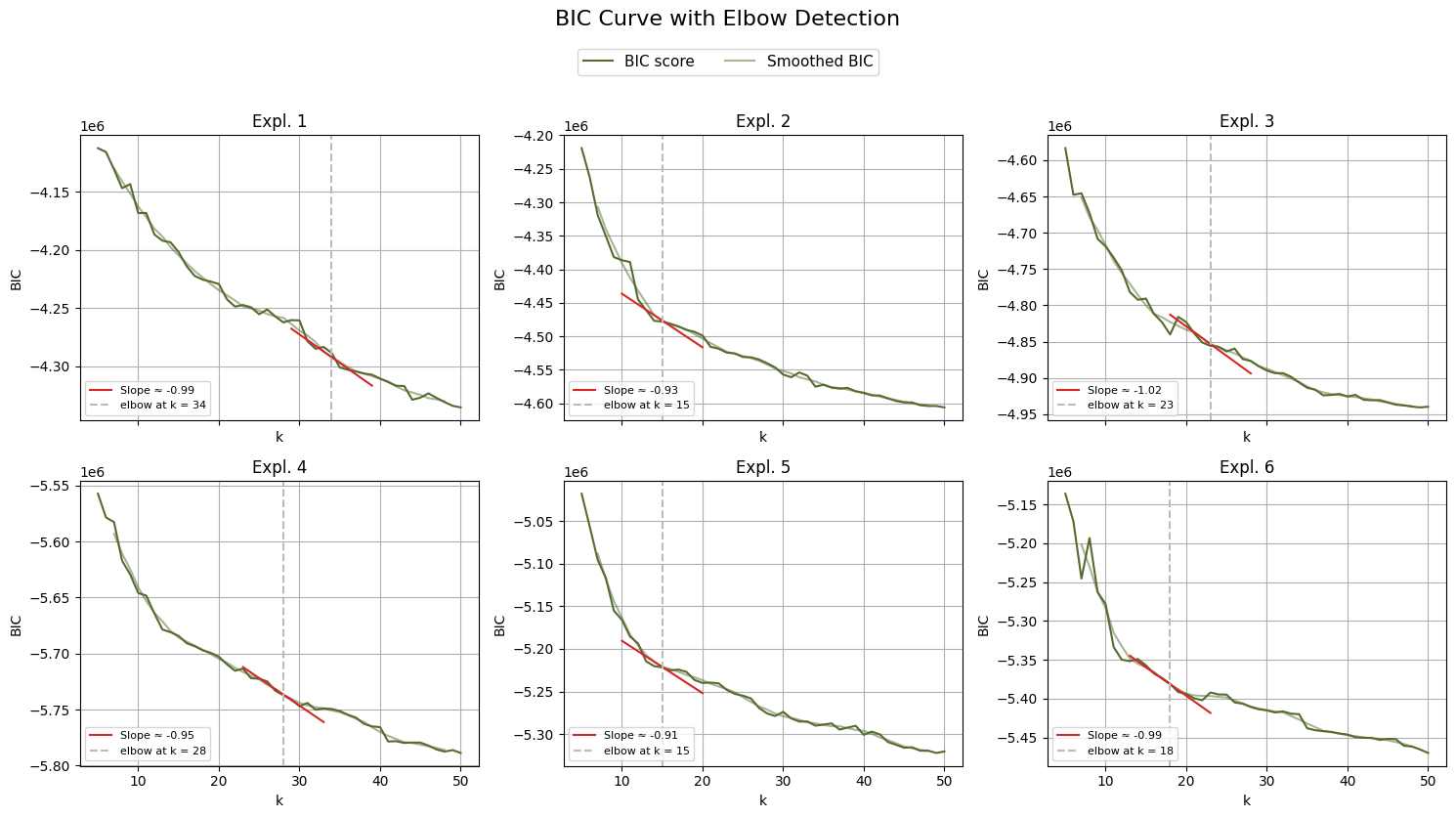}
    \caption{Smoothed BIC curves and estimated elbow points for each exploration. The dashed line indicates the elbow (i.e., where the slope of the tangent first crosses \(-1\)), and the red segment corresponds to the portion of the curve used for slope estimation.}
    \label{fig:bic_gmm}
\end{figure}

Figure~\ref{fig:bic_gmm} shows the \ac{BIC} curves obtained for each exploration as a function of $k$. The \ac{BIC} score was smoothed to reduce noise, and the elbow was estimated as the point where the slope of the tangent line approached \(-1\). Although this method yielded plausible estimates of $k$ in some cases, it failed to provide consistent or informative results across all explorations.

\subsection{Model Hyperparameters and Performance}
\label{sec:appendix_hyperparams}

This section reports the best-performing configurations for the classification models evaluated in this work. 

Two strategies were compared for selecting the number of \acp{HDA} \(k\): (1) fixing \(k\) to the elbow point of the \ac{BIC} curve; and, (2) an exhaustive search of \(k\) via cross-validation. In both cases, the classifier and its hyperparameters were selected based on validation performance. Interestingly, when \(k\) was fixed via \ac{BIC}, \ac{RF} performed best across all explorations. When \(k\) was optimised during model selection, the \ac{SVM}-RBF consistently outperformed other classifiers.

\subsubsection{\ac{RF} with \ac{BIC}-Based Selection of \(k\)}

Table~\ref{tab:bic_rf_results} reports the best-performing configurations for each exploration when the number of \acp{HDA} \(k\) was fixed using the \ac{BIC} elbow method, trained in EF3. For each setting, the classifier type and hyperparameters were selected via grid search using cross-validated \ac{AUC}. \acp{RF} emerged as the best-performing model under this constraint.

When the elbow-based values of \(k\) derived from the BIC curves were used in the classification pipeline, the resulting models exhibited lower performance, particularly in terms of average \ac{AUC}. As shown in Table~\ref{tab:bic_rf_results}, the best-performing configuration under this strategy reached an AUC of \(0.75 \pm 0.15\), while most others remained below 0.65. In contrast, the SVM-RBF models with tuned \(k\) (Table~\ref{tab:svm_results}) achieved AUCs as high as \(0.81 \pm 0.12\). As a result, the BIC-based strategy was not retained in the final framework.

\begin{table}[h!]
\centering
\caption{Best configurations for \ac{RF} classifiers using \ac{BIC}-derived \(k\) values.}
\label{tab:bic_rf_results}
\begin{tabular}{|c|c|c|c|}
\hline
\textbf{Exploration} & \textbf{$k$} & \textbf{Max Depth} & \textbf{Average \ac{AUC} (\(\pm\) std)} \\ \hline
\textit{Expl.~1} & 34 & 12 & \(0.67 \pm 0.12\) \\
\textit{Expl.~2} & 15 & 5  & \(0.75 \pm 0.15\) \\
\textit{Expl.~3} & 23 & 4  & \(0.62 \pm 0.12\) \\
\textit{Expl.~4} & 28 & 1  & \(0.64 \pm 0.11\) \\
\textit{Expl.~5} & 15 & 1  & \(0.63 \pm 0.10\) \\
\textit{Expl.~6} & 18 & 1  & \(0.60 \pm 0.16\) \\ \hline
\end{tabular}
\end{table}

\subsubsection{\ac{SVM}-RBF with Optimised \(k\)}

The \ac{SVM}-RBF classifiers trained under EF3 were first optimised by jointly tuning the regularisation parameter \(C\), the kernel coefficient \(\gamma\), and the number of components \(k\), using cross-validation. Table~\ref{tab:svm_optimised_results} presents the best performing configurations obtained for each exploration under this fully optimised setting.

To reduce model complexity and encourage generalisability, a simplified version of the model was also evaluated, in which the kernel coefficient was fixed at \(\gamma = 0.01\) for all explorations. This decision was motivated by the fact that \(\gamma = 0.01\) emerged as the optimal value in four out of the six explorations during the initial hyperparameter search. Fixing \(\gamma\) reduces the number of parameters to be tuned and mitigates overfitting to exploration-specific idiosyncrasies. The performance drop observed with this simplification was minimal: across explorations, average \ac{AUC} values changed by less than 0.02 in absolute terms. The resulting configurations with fixed \(\gamma\) are shown in Table~\ref{tab:svm_results}.

\begin{table}[ht]
\centering
\caption{Best configurations for \ac{SVM}-RBF classifiers (EF3), with all parameters optimised independently for each exploration.}
\label{tab:svm_optimised_results}
\begin{tabular}{|c|c|c|c|c|}
\hline
\textbf{Exploration} & \textbf{$k$} & \textbf{$C$} & \textbf{$\gamma$} & \textbf{Average \ac{AUC} (\(\pm\) std)} \\ \hline
\textit{Expl.~1} & 6  & 1.0   & 0.01  & \(0.75 \pm 0.10\) \\
\textit{Expl.~2} & 23 & 1.0   & 0.01  & \(0.81 \pm 0.11\) \\
\textit{Expl.~3} & 23 & 100.0 & 0.001 & \(0.61 \pm 0.10\) \\
\textit{Expl.~4} & 8  & 10.0  & 0.001 & \(0.71 \pm 0.12\) \\
\textit{Expl.~5} & 44 & 10.0  & 0.01  & \(0.66 \pm 0.10\) \\
\textit{Expl.~6} & 10 & 10.0  & 0.01  & \(0.72 \pm 0.14\) \\ \hline
\end{tabular}
\end{table}

\begin{table}[ht]
\centering
\caption{Best configurations for \ac{SVM}-RBF classifiers (EF3), with \(\gamma = 0.01\) fixed across explorations.}
\label{tab:svm_results}
\begin{tabular}{|c|c|c|c|}
\hline
\textbf{Exploration} & \textbf{$k$} & \textbf{$C$} & \textbf{Average \ac{AUC} (\(\pm\) std)} \\ \hline
\textit{Expl.~1} & 6  & 1.0   & \(0.75 \pm 0.10\) \\
\textit{Expl.~2} & 23 & 1.0   & \(0.81 \pm 0.12\) \\
\textit{Expl.~3} & 48 & 1.0   & \(0.61 \pm 0.11\) \\
\textit{Expl.~4} & 8  & 1.0   & \(0.70 \pm 0.12\) \\
\textit{Expl.~5} & 44 & 100.0 & \(0.66 \pm 0.10\) \\
\textit{Expl.~6} & 10 & 10.0  & \(0.72 \pm 0.14\) \\ \hline
\end{tabular}
\end{table}

\newpage
\bibliographystyle{unsrt}
\bibliography{references}

\begin{thebibliography}{10}

\bibitem{reiner2023oculometric}
Johnathan Reiner, Liron Franken, Eitan Raveh, Israel Rosset, Rivka Kreitman, Edmund Ben-Ami, and Ruth Djaldetti.
\newblock Oculometric measures as a tool for assessment of clinical symptoms and severity of parkinson’s disease.
\newblock {\em Journal of Neural Transmission}, 130:1241--1248, 2023.

\bibitem{matsumoto2011small}
Hideaki Matsumoto, Takashi Hanakawa, Tianzi Wu, Kenji Kansaku, and Mark Hallett.
\newblock Small saccades restrict visual scanning area in parkinson’s disease.
\newblock {\em Movement Disorders}, 26(9):1619--1626, 2011.

\bibitem{bek2020measuring}
Judith Bek, Ellen Poliakoff, and Karen Lander.
\newblock Measuring emotion recognition by people with parkinson’s disease using eye-tracking with dynamic facial expressions.
\newblock {\em Journal of Neuroscience Methods}, 331:108524, 2020.

\bibitem{tsang2016eye}
Vicky Tsang.
\newblock Eye-tracking study on facial emotion recognition tasks in individuals with high-functioning autism spectrum disorders.
\newblock {\em Autism}, 22(2):161--170, 2016.

\bibitem{armstrong2012eye}
Thomas Armstrong and Bunmi~O. Olatunji.
\newblock Eye tracking of attention in the affective disorders: A meta-analytic review and synthesis.
\newblock {\em Clinical Psychology Review}, 32(8):704--723, 2012.

\bibitem{davis2020eye}
Rebecca Davis and Alla Sikorskii.
\newblock Eye tracking analysis of visual cues during wayfinding in early stage alzheimer’s disease.
\newblock {\em Dementia and Geriatric Cognitive Disorders}, 49:91--97, 2020.

\bibitem{boz2023examination}
Hatice~Eraslan Boz, Koray Kocogglu, Muge Akkoyun, Isil~Yaggmur Tufekci, Merve Ekin, Pinar Ozcelik, and Gulden Akdal.
\newblock Examination of eye movements during visual scanning of real-world images in alzheimer’s disease and amnestic mild cognitive impairment.
\newblock {\em International Journal of Psychophysiology}, 190:84--93, 2023.

\bibitem{metternich2022eye}
Birgitta Metternich, Nina~A. Gehrer, Kathrin Wagner, Maximilian~J. Geiger, Elisa Schütz, Andreas Schulze-Bonhage, Marcel Heers, and Michael Schönenberg.
\newblock Eye-movement patterns during emotion recognition in focal epilepsy: An exploratory investigation.
\newblock {\em Seizure: European Journal of Epilepsy}, 100:95--102, 2022.

\bibitem{ashaie2020eye}
Sameer~A. Ashaie and Leora~R. Cherney.
\newblock Eye tracking as a tool to identify mood in aphasia: A feasibility study.
\newblock {\em Neurorehabilitation and Neural Repair}, 34(5):463--471, 2020.

\bibitem{wang2015atypical}
Shuo Wang, Ming Jiang, Xavier~Morin Duchesne, Elizabeth~A. Laugeson, Daniel~P. Kennedy, Ralph Adolphs, and Qi~Zhao.
\newblock Atypical visual saliency in autism spectrum disorder quantified through model-based eye tracking.
\newblock {\em Neuron}, 88(3):604--616, 2015.

\bibitem{antoniades2024eye}
Chrystalina~A. Antoniades and Miriam Spering.
\newblock Eye movements in parkinson’s disease: From neurophysiological mechanisms to diagnostic tools.
\newblock {\em Trends in Neurosciences}, 47(1):71--80, 2024.

\bibitem{kassavetis2022eye}
Panagiotis Kassavetis, Diego Kaski, Tim Anderson, and Mark Hallett.
\newblock Eye movement disorders in movement disorders.
\newblock {\em Movement Disorders Clinical Practice}, 9(3):284--295, 2022.

\bibitem{rodriguez2019eye}
Roberto Rodríguez-Labrada, Yaimeé Vázquez-Mojena, and Luis Velázquez-Pérez.
\newblock {\em Eye Movement Abnormalities in Neurodegenerative Diseases}.
\newblock IntechOpen, 2019.

\bibitem{leigh2020abnormal}
R.~John Leigh and David~S. Zee.
\newblock Abnormal eye movements in parkinsonism: A historical view.
\newblock {\em Movement Disorders}, 2020.

\bibitem{wong2020prolonged}
Wing~Ho Wong, Vincent Mok, Anne Chan, Adrian Wong, and Sandra~SM Chan.
\newblock Prolonged visual fixation as a surrogate marker of cholinergic deficit in parkinson’s disease.
\newblock {\em Parkinsonism and Related Disorders}, 81:60--66, 2020.

\bibitem{dietz2011emotion}
J.~Dietz, M.M. Bradley, M.S. Okun, and D.~Bowers.
\newblock Emotion and ocular responses in parkinson’s disease.
\newblock {\em Neuropsychologia}, 49(12):3247--3253, 2011.

\bibitem{wong2018eye}
Oscar~WH Wong, Anne~YY Chan, Adrian Wong, Claire~KY Lau, Jonas~HM Yeung, Vincent~CT Mok, Linda~CW Lam, and Sandra Chan.
\newblock Eye movement parameters and cognitive functions in parkinson’s disease patients without dementia.
\newblock {\em Parkinsonism and Related Disorders}, 52:43--48, 2018.

\bibitem{takemoto2023depression}
Ayumi Takemoto, Inese Aispuriete, Laima Niedra, and Lana~Franceska Dreimane.
\newblock Depression detection using virtual avatar communication and eye tracking.
\newblock {\em Journal of Eye Movement Research}, 16(2):6, 2023.

\bibitem{arndt2014eye}
Jody~E. Arndt, Kristin~R. Newman, and Christopher~R. Sears.
\newblock An eye tracking study of the time course of attention to positive and negative images in dysphoric and non-dysphoric individuals.
\newblock {\em Journal of Experimental Psychopathology}, 5(4):399--413, 2014.

\bibitem{russell2015eye}
Kristin Russell and Christopher~Roy Sears.
\newblock Eye gaze tracking reveals different effects of a sad mood induction on the attention of previously depressed and never depressed women.
\newblock {\em Cognitive Therapy and Research}, 39:292--306, 2015.

\bibitem{coutrot2018scanpath}
Antoine Coutrot, Janet~H. Hsiao, and Antoni~B. Chan.
\newblock Scanpath modeling and classification with hidden markov models.
\newblock {\em Behavior Research Methods}, 50:362--379, 2018.

\bibitem{rabiner1989tutorial}
Lawrence~R Rabiner.
\newblock A tutorial on hidden markov models and selected applications in speech recognition.
\newblock {\em Proceedings of the IEEE}, 77(2):257--286, 1989.

\bibitem{tibon2021transient}
Roni Tibon, Kamen~A. Tsvetanov, Darren Price, David Nesbitt, and Richard Henson.
\newblock Transient neural network dynamics in cognitive ageing.
\newblock {\em Neurobiology of Aging}, 105:217--228, 2021.

\bibitem{bustamante2023classification}
Catalina Bustamante, Gabriel Castrill{\'o}n, and Juli{\'a}n Arias-Londo{\~n}o.
\newblock Classification of focused perturbations using time-variant functional connectivity with rs-fmri.
\newblock In {\em Colombian Conference on Computing (ColCACI)}, volume 1746 of {\em Communications in Computer and Information Science (CCIS)}, pages 18--30. Springer, Springer, Cham, 2023.

\bibitem{khorasani2014hmm}
Abed Khorasani and Mohammad~Reza Daliri.
\newblock Hmm for classification of parkinson’s disease based on the raw gait data.
\newblock {\em Journal of Medical Systems}, 38(12):147, 2014.

\bibitem{arias2015entropies}
Julián~D. Arias-Londoño and Juan~I. Godino-Llorente.
\newblock Entropies from markov models as complexity measures of embedded attractors.
\newblock {\em Entropy}, 17(6):3595--3620, 2015.

\bibitem{heideman2020dissecting}
Simone~G. Heideman, Andrew~J. Quinn, Mark~W. Woolrich, Freek van Ede, and Anna~C. Nobre.
\newblock Dissecting beta-state changes during timed movement preparation in parkinson’s disease.
\newblock {\em Progress in Neurobiology}, 184:101731, 2020.

\bibitem{zhu2023mibfm}
Jing Zhu, Changlin Yang, Xiannian Xie, Shiqing Wei, Yizhou Li, Xiaowei Li, and Bin Hu.
\newblock Mutual information based fusion model (mibfm): Mild depression recognition using eeg and pupil area signals.
\newblock {\em IEEE Transactions on Affective Computing}, 14(3):2102--2111, 2023.

\bibitem{coutrot2017scanpath}
Antoine Coutrot, Janet~H. Hsiao, and Antoni~B. Chan.
\newblock Scanpath modeling and classification with hidden markov models.
\newblock {\em Behavior Research Methods}, 50(1):362--379, 2018.

\bibitem{arias2010automatic}
Juli{\'a}n~D Arias-Londono, Juan~I Godino-Llorente, Nicol{\'a}s S{\'a}enz-Lech{\'o}n, V{\'\i}ctor Osma-Ruiz, and Germ{\'a}n Castellanos-Dom{\'\i}nguez.
\newblock Automatic detection of pathological voices using complexity measures, noise parameters, and mel-cepstral coefficients.
\newblock {\em IEEE Transactions on biomedical engineering}, 58(2):370--379, 2010.

\bibitem{osterrieth1944test}
P.~A. Osterrieth.
\newblock Le test de copie d’une figure complexe; contribution à l’étude de la perception et de la mémoire. [test of copying a complex figure; contribution to the study of perception and memory.].
\newblock {\em Archives de Psychologie}, 30:206--356, 1944.

\bibitem{srresearch2017eyelink}
SR~Research Ltd.
\newblock {\em EyeLink 1000 Plus User Manual}.
\newblock SR Research Ltd., 2017.
\newblock Version 1.0.12.

\bibitem{bishop2006pattern}
Christopher~M. Bishop.
\newblock {\em Pattern Recognition and Machine Learning}.
\newblock Springer, 2006.

\bibitem{shannon1948}
Claude~E. Shannon.
\newblock A mathematical theory of communication.
\newblock {\em Bell System Technical Journal}, 27(3):379--423, 1948.

\bibitem{schwarz1978bic}
Gideon~E. Schwarz.
\newblock Estimating the dimension of a model.
\newblock {\em Annals of Statistics}, 6(2):461--464, 1978.

\bibitem{hastie2009elements}
Trevor Hastie, Robert Tibshirani, and Jerome Friedman.
\newblock {\em The Elements of Statistical Learning: Data Mining, Inference, and Prediction}.
\newblock Springer, 2nd edition, 2009.

\bibitem{sharkey1996combining}
Amanda~J.C. Sharkey.
\newblock On combining artificial neural nets.
\newblock {\em Connection Science}, 8(3-4):299--314, 1996.

\bibitem{breiman1996stacked}
Leo Breiman.
\newblock Stacked regressions.
\newblock {\em Machine Learning}, 24(1):49--64, 1996.

\bibitem{kittler1998combining}
Josef Kittler, M~Hatef, RPW Duin, and J~Matas.
\newblock On combining classifiers.
\newblock {\em IEEE Transactions on Pattern Analysis and Machine Intelligence}, 20(3):226--239, 1998.

\bibitem{zhou2012ensemble}
Zhi-Hua Zhou.
\newblock {\em Ensemble Methods: Foundations and Algorithms}.
\newblock CRC Press, 2012.

\bibitem{borji2013state}
Ali Borji and Laurent Itti.
\newblock State-of-the-art in visual attention modeling.
\newblock {\em IEEE transactions on pattern analysis and machine intelligence}, 35(1):185--207, 2013.

\bibitem{luque2024estimation}
Elisa Luque-Buzo, Mehdi Bejani, Juli{\'a}n~D Arias-Londo{\~n}, Jorge~A G{\'o}mez-Garc{\'\i}a, Francisco Grandas-P{\'e}rez, and Juan~I Godino-Llorente.
\newblock Estimation of the cyclopean eye from binocular smooth pursuit tests.
\newblock {\em IEEE Transactions on Cognitive and Developmental Systems}, 16(6):2125--2137, 2024.

\bibitem{dowiasch2015aging}
Stefan Dowiasch, Sebastian Marx, Wolfgang Einh{\"a}user, and Frank Bremmer.
\newblock Effects of aging on eye movements in the real world.
\newblock {\em Frontiers in Human Neuroscience}, 9:46, 2015.

\end{thebibliography}

\end{document}